\documentclass[%
 reprint,
 amsmath,amssymb,
 aps,
]{revtex4-1}

\usepackage{graphicx}
\usepackage{epspdfconversion}
\usepackage{dcolumn}
\usepackage{bm}

\begin{document}

\preprint{APS/123-QED}

\title{Fickian yet non-Gaussian diffusion in two-dimensional Yukawa liquids}

\author{Zahra Ghannad}
 \altaffiliation{z.ghannad@alzahra.ac.ir}

\affiliation{%
 Department of Physics, Alzahra University, P. O. Box 19938-93973, Tehran, Iran
}%

\date{\today}

\begin{abstract}
We investigate Fickian diffusion in two-dimensional (2D) Yukawa liquids using molecular dynamics simulations. We compute the self-van Hove correlation function $G_s(r,t)$, and self-intermediate scattering function $F_s(k,t)$ and compare these functions with those obtained from mean-squared displacement  MSD using the Gaussian approximation. According to this approximation, a linear MSD with time implies a Gaussian behavior for $G_s(r,t)$  and $F_s(k,t)$ at all times. Surprisingly, we find that these functions deviate from Gaussian at intermediate time scales, indicating the failure of the Gaussian approximation. Furthermore, we quantify these deviations by the non-Gaussian parameter, and we find that the deviations increase with decreasing the temperature of the liquid. The origin of the non-Gaussian behavior may be the heterogeneous dynamics of dust particles observed in 2D Yukawa liquids.
\begin{description}
\item[PACS numbers]
52.27.Lw, 52.27.Gr
\item[Keywords]
Fickian diffusion, Heterogeneous dynamics, Non-Gaussian parameter, Yukawa liquid, Self-van Hove function

\end{description}
\end{abstract}

\maketitle


\section{\label{sec:level1}Introduction}

Dusty plasma is a weakly ionized gas containing ions, electrons, neutrals, and highly charged dust particles.~\cite{Bonitz2010,Fortov2005,Morfill2009,Khrapak2018,Feng2013}. In the laboratory, due to the electric field in the plasma sheath, dust particles can be levitated and confined. Therefore, they can be floated in a monolayer, with an ignorable out-of-plan motion to form a two dimensional (2D) dusty plasma~\cite{Feng2016,Wang2018,Feng2017}. Because of shielding by electrons and ions of the background plasma, the interaction potential between dust particles is accurately described by Yukawa potential, i.e., $\phi(r)= Q^2\rm{exp}(-\it{r}/\lambda_D)/\rm{4}\pi \epsilon_0\it{r}$, where $\lambda_D$ is the Debye shielding length and $Q$ is the dust charge ~\cite{Yukawa1935,Konpka2000}. The high charges of dust particles cause their electrostatic potential energy to exceed the kinetic energy. Hence, dust particles are strongly coupled so that the collection of them exhibits behaviors of liquids~\cite{Feng2016,Wang2018,Feng2017} and solids~\cite{Feng2008,Hartmann2014}.
Here, we focus on the Brownian diffusion of dust particles in 2D equilibrium Yukawa liquids. 

The random motion of dust particles in 2D Yukawa liquids can be described by the Brownian diffusion~\cite{Liu2006}.
It was observed by Brown~\cite{Brown1828}, and its theoretical description was derived by Einstein~\cite{Einstein1905}. There are two fundamental features with Brownian diffusions~\cite{Chechkin2017}:

(i) The mean-squared displacement $\langle\Delta \rm{r}^2(t)\rangle$, is linear with time
\begin{equation}
\langle(\Delta \mathrm{r}(t))^2\rangle=\langle\vert\boldsymbol{\mathrm{r}}(t)-\boldsymbol{\mathrm{r}}(0)\vert^2\rangle=\int \mathrm{r}^2 G_s(\boldsymbol{\mathrm{r}},t)d\boldsymbol{\mathrm{r}}=2dDt,\label{eq1}
\end{equation}
which is called Fickian (normal) diffusion. Here, $D$ is the diffusion coefficient and $d$ denotes the dimension.

(ii) The self-part of the van Hove correlation function, i.e., the distribution of the particle displacement is Gaussian
\begin{equation}
 G_s(\boldsymbol{\mathrm{r}},t)= \frac{1}{(4\pi Dt)^{d/2}}\mathrm{exp}\left(-\frac{\boldsymbol{\mathrm{r}}^2}{4Dt}\right).\label{eq2}
\end{equation}
$ G_s(\boldsymbol{\mathrm{r}},t)d\boldsymbol{\mathrm{r}}$ gives the probability of finding a particle at position   \boldsymbol{\mathrm{r}} at time $t$ given that the same particle was at the origin at the initial time $t=0$~\cite{vanHove1954,Hansen,Hail}.

In this work, by molecular dynamics (MD) simulation, we investigate the diffusion process in 2D equilibrium Yukawa liquids. Investigating the diffusion in 2D Yukawa liquids has attracted a great deal of interest over the last decade~\cite{Liu2007,Ott2008,Hartmann2019,Feng2010,Tau1998,Vaulina2006,Donko2009,Nunomura2006,Hou2009,Feng2014,
Tau2001,Dzhumagulova2014}. We find that the distribution function of dust particles displacement $G_s(r,t)$ is not Gaussian as expected for a Fickian diffusion while the mean-squared displacement appears Fickian (linear with time). Fickian yet non-Gaussian diffusions have also been observed in various systems, such as colloids~\cite{Kown2014,Schnyder2017,Kim2013,Guan2014} and porous media~\cite{He2013,He2014}.

The paper is organized as follows. In Section \ref{Model}, we describe the fundamental features of the simulation technique to mimic 2D Yukawa liquids. In Section \ref{Results}, we compute four diagnostics including the mean-squared displacement, the self-part of the van Hove correlation function, the non-Gaussian parameter, and the self-part of the intermediate scattering function to interpret underlying physics. In Section \ref{Conclusions}, we present the conclusions.

\section{\label{Model}MODEL AND SIMULATION TECHNIQUE}
To study 2D equilibrium Yukawa liquids, we performed equilibrium MD simulations~\cite{Frenkel}. We integrated the equation of the motion $m\ddot{\boldsymbol{\mathrm{r}}}_i=-\nabla\Sigma_j\phi_{ij}$ for $N$= 1024 particles, where $\phi_{ij}$ is the Yukawa pair interaction potential~\cite{Yukawa1935,Konpka2000}. An equilibrium Yukawa system can be characterized by two dimensionless parameters~\cite{Ohta2000,Hartmann2019}: 

$\bullet$ The Coulomb coupling parameter $\Gamma=Q^2/4\pi\epsilon_0ak_BT$, where $\epsilon_0$ is the dielectric constant, $k_B$ is the Boltzmann constant, $Q$ is the charge of a dust particle, $T$ is the kinetic temperature of dust particles, $a=(n\pi)^{-1/2}$ is the Wigner-Seitz radius for 2D systems, and $n$ is the surface number density of dust particles.

$\bullet$ The screening parameter $\kappa=a/\lambda_D$.\\ 
 The dimensionless units in this work are listed in  Table~\ref{jlab1}, where $\omega_{pd}=(Q^2/2\pi \epsilon_0 ma^3)^{1/2}$ is the nominal 2D dusty plasma frequency~\cite{Kalman2004}.
\begin{table}[ht]
\caption{\label{jlab1} Dimensionless units for 2D Yukawa liquids.}
\begin{ruledtabular}
\bgroup
\def\arraystretch{1.4}
\begin{tabular}{cccccccc}
Quantity&Symbol&&Dimensionless unit \\
\hline
Length&$r$&$\longrightarrow$& $r/a$\\
Time&$t$&$\longrightarrow$&$\omega_{pd}t$\\
Temperature&$T$&$\longrightarrow$&$\Gamma^{-1}$\\
Surface number density&$n$&$\longrightarrow$&$na^2$\\
Potential energy&$\phi$&$\longrightarrow$&$\phi/a^2\omega_{pd}^2$\\
Wave number&$k$&$\longrightarrow$&$ka$\\
\end{tabular}
\egroup
\end{ruledtabular}
\end{table}

The values of $\Gamma$ and $\kappa$  are entered as input parameters in simulations. We chose three values of $\kappa =$  0.5, 1.2, 2.0 as the beginning, the middle and the end of the allowed interval
from an experimentally relevant range of 0.5 $\leqslant$$\kappa$ $\leqslant$ 2.0~\cite{Donko2006}. For each $\kappa$ value, there is one specific melting point $\Gamma_m$, so that for $\Gamma$ $<$ $\Gamma_m$, the 2D Yukawa system is in the liquid phase as reported by Hartmann $\textit{et al.}$~\cite{Hartmann2005}. These values are $\Gamma_m=$ 142 for $\kappa=$ 0.5~\cite{Feng2013,Hartmann2005}, $\Gamma_m=$ 200 for $\kappa=$ 1.2~\cite{Hartmann2005}, $\Gamma_m =$ 415 for $\kappa=$ 2.0~\cite{Ott2008,Hartmann2005}. 
In simulations, we chose the values of $\Gamma$ over a range that allows simulations of liquids.

The simulation box is a rectangular box with the size of 61.1$a\times52.9a$, so that the surface number density is consistent with the definition of the Wigner-Seitz radius, i.e., $n= 1024/( 61.1a\times 52.9a)\approx1/(\pi a^2)$~\cite{Fengpop2016}. To eliminate boundary effects caused by the finite size of the simulation box, and model the system as an infinite one, we applied the periodic boundary conditions.

We began from an initial random configuration of dust particles and used a Nos$\acute{e}$-Hoover thermostat~\cite{Nose1984,Hoover1985} to reach the system at the desired temperature. Then, we turned off the thermostat to sample dynamical properties. 

We used the velocity Verlet algorithm~\cite{Swope1982} to integrate the equations of the motion with the integration time step of 0.037 $\omega^{-1}_{pd}$, which we verified that this time step is adequately small to conserve energy. Generally, the time step is chosen from the range between (0.0037--0.037) $\omega^{-1}_{pd}$ depending on the $\Gamma$ values; i.e., 0.0037 $\omega^{-1}_{pd}$ for 1.0 $\leqslant$ $\Gamma$ $<$ 4.0, 0.0093 $\omega^{-1}_{pd}$ for 4.0 $\leqslant$ $\Gamma$ $<$ 10.0, 0.0185 $\omega^{-1}_{pd}$ for 10.0 $\leqslant$ $\Gamma$ $<$ 40.0, and 0.037 $\omega^{-1}_{pd}$ for $\Gamma$ $\geqslant$ 40.0 ~\cite{Feng2016}. The last one is chosen in our simulations because we studied $\Gamma$ $>$ 100.	

 Since the Yukawa potential decays as exp(-$r$)$/r$, the cutoff radius for this potential in MD simulations should be sufficiently large to ensure that the perturbation introduced into simulations due to the potential truncation is negligible~\cite{Bonitz2010}. We truncated the Yukawa potential at $r_{\rm{cut}}$ = 24.8$a$, as in~\cite{Feng2016,Wang2018,Feng2017}.

To verify that our MD simulations reasonably modeled a canonical ensemble in thermal equilibrium, we applied a standard test as follows:\\
In a finite system in equilibrium, the temperature fluctuates about the mean value, i.e., $\delta T= T-\langle T\rangle$. If the system exhibits canonical fluctuations (within a computationally reasonable time), then, the variance of the temperature is~\cite{Holian1995}
\begin{equation}
\langle(\delta T)^2\rangle=\frac{2}{d}\frac{\langle T\rangle^2}{N},
\label{eq3}
\end{equation} 
where $\langle T\rangle$ is the mean temperature. We calculated the variance of the temperature in our simulations and compared it to the variance for the canonical ensemble in thermal equilibrium, given in the above equation. According to the standard test, the ratio of these two variances is unity for a canonical system in thermal equilibrium. Although a value of unity is an ideal value, in a simulation due to a limited time range for sampling, a value very close to 1 is considered successful. We found that the ratio was 0.996, which assured us our simulations precisely modeled a canonical system in equilibrium.

\section{\label{Results}RESULTS AND DISCUSSIONS}
In this section, we use four diagnostics to characterize the dynamics of dust particles in 2D Yukawa liquids. They are the mean-squared displacement, the self-part of the van Hove correlation function,  the non-Gaussian parameter, and the self-intermediate scattering function, as explained below.

\subsection{\label{MSD}Mean-squared displacement}
The first diagnostic, mean-squared displacement (MSD), $\langle(\Delta \mathrm{r}(t))^2\rangle = \langle\vert\boldsymbol{\mathrm{r}}_i(t)-\boldsymbol{\mathrm{r}}_i(0)\vert^2\rangle$ is calculated to identify Fickian (normal) diffusion. Here, $\boldsymbol{\mathrm{r}}_i(t)$ is the position of particle $i$ at time $t$ and $\langle...\rangle$ denotes an ensemble average.
For a 2D system, the MSD obeys a power law MSD($t$) $\propto$ 4$Dt^{\alpha}$. When MSD plot as a function of time in a log-log plot, the fitted curve with a straight line gives the slope $\alpha$. The
signatures of normal diffusion and anomalous diffusion are $\alpha$ = 1
and $\alpha\neq$ 1, respectively. Anomalous diffusion refers to both superdiffusion ($\alpha>$ 1) and subdiffusion ($\alpha<$ 1). Since data from simulations will never yield a value that is exactly 1, a range of $0.9<\alpha<1.1$ is classified as normal diffusion, and $\alpha>1.1$ for superdiffusion~\cite{Feng2014,Liu2007,Feder1996}.

Results for MSD with selected values of $\kappa$ and $\Gamma$ are shown as a function of time in a log-log scale in Figs. 1(a)--1(c). At very short times ($\omega_{pd}t$ $\lesssim$ 5) when dust particles move in the cage created by neighboring dust particles, all curves are fitted by a straight line with a slope $\alpha =$ 2, indicating the ballistic motion. At later times, dust particles escape from the cages and diffuse. Here we are interested in motion at intermediate times (when the Gaussian approximation fails). For a period of $100<\omega_{pd}t<600$, the fitting results of the exponent $\alpha$  for $\kappa=0.5$ are $\alpha=1.15$ for $\Gamma=100$, $\alpha=1.08$ for $\Gamma=120$, and $\alpha=1.05$ for $\Gamma=130$.  These results for $\kappa=1.2$ are  $\alpha=1.15$ for $\Gamma=100$, $\alpha=1.11$ for $\Gamma=140$, and $\alpha=1.05$ for $\Gamma=190$. 
For $\kappa=2.0$, we obtain $\alpha=1.09$ for $\Gamma=240$, $\alpha=1.07$ for $\Gamma=320$, and $\alpha=1.04$ for $\Gamma=400$. All these results indicate with increasing $\Gamma$ (equivalently decreasing temperature) to near $\Gamma_m$ ($T_m$), superdiffusion of dust particles tends to normal diffusion. Therefore, for very low temperatures, i.e., supercooled Yukawa liquids ($\Gamma>120$ for $\kappa=0.5$, $\Gamma>140$ for $\kappa=1.2$, $\Gamma>320$ for $\kappa=2.0$ ), the diffusion of dust particles is normal (Fickian) with good accuracy.

\subsection{\label{vanHove}Self-part of the van Hove correlation function}
 The self-part of the van Hove correlation function, i.e., the distribution of the particle displacement is defined as~\cite{Hail}
\begin{equation}
G_s(\boldsymbol{\mathrm{r}},t)=\frac{1}{N}\Bigg\langle\sum_{\mathit{i}=1}^N\delta(\boldsymbol{\mathrm{r}}-\boldsymbol{\mathrm{r}}_i(t)+\boldsymbol{\mathrm{r}}_i(0))\Bigg\rangle,
\label{eq4}
\end{equation}
where $\langle...\rangle$ represents an ensemble average and $\delta$ is the Dirac delta function. 
For isotropic liquids, $G_s$ depends only on the scalar  
distance, $r=\vert\boldsymbol{\mathrm{r}}\vert$. Thus, Eq.(\ref{eq4}) reduces to
\begin{equation}
G_s(r,t)=\frac{1}{N}\Bigg\langle\sum_{\mathit{i}=1}^N\delta(r-\vert r_i(t)-r_i(0)\vert)\Bigg\rangle
\label{eq5}.
\end{equation}
Physically, 2$\pi rG_s(r,t)dr$ measures the probability of finding a dust particle at distance $r$ 
from an origin at time $t$ given that the same dust particle was at the origin at the initial time $t=0$~\cite{Hail}. Thus, $G_s(r,t)$ is normalized by
\begin{equation}
\int G_s(r,t)d\boldsymbol{\mathrm{r}}=1
\label{eq6}.
\end{equation}

Figures 2(a)--2(c), 4(a)--4(c), and 6(a)--6(c) show the time evolution of the normalized $G_s(r,t)$ for the selected $\kappa$ and $\Gamma$ values, and compare $G_s\rm{(}\it{r},t\rm{)}$ with Gaussian distribution $G_s^{\rm{Gauss}}\rm{(}\it{r},t\rm{)}$ (solid lines) obtained from Eq.~(\ref{eq2}), i.e., $G_s^{\rm{Gauss}}\rm{(}\it{r},t\rm{)}=(1/\pi\langle(\Delta \mathrm{r}(t))^2\rangle)\rm{exp}(-r^2/\langle(\Delta \mathrm{r}(t))^2\rangle)$, where $\langle(\Delta \mathrm{r}(t))^2\rangle$ is obtained from the simulation.
At intermediate times, when the MSD is linear with time, $G_s(r,t)$ functions deviate from the Gaussian.

How much difference is between $G_s\rm{(}\it{r},t\rm{)}$ and the Gaussian distribution? To answer this question, we calculate the relative difference between $G_s\rm{(}\it{r},t\rm{)}$ and $G_s^{\rm{Gauss}}\rm{(}\it{r},t\rm{)}$, i.e., $G_s\rm{(}\it{r},t\rm{)}$-$G_s^{\rm{Gauss}}\rm{(}\it{r},t\rm{)}$/ $G_s^{\rm{Gauss}}\rm{(}\it{r},t\rm{)}$ as shown in Figs. 3(a)--3(c), 5(a)--5(c), and 7(a)--7(c).
 Maximum relative differences between $G_s(r,t)$ and $G_s^{Gauss}(r,t)$ become as large as 10, $10^3,10^4$ for $\kappa =2, \Gamma=$ 240, 320, 400 respectively, and 10, $10^2 ,10^3$ for $\kappa=1.2, \Gamma=$ 100, 140, 190, and  10, $10^2,10^2$ for $\kappa =0.5, \Gamma=$ 100, 120, 130 respectively.

 For a fixed value of $\kappa$, the maximum relative difference increases with increasing $\Gamma$, i.e., with decreasing $T$, and
it  is as large as $10^4$, which belongs to the lowest temperature, i.e., $T$= 0.0025 or $\Gamma=400$ ( see figure 3(c)).
 It means that in the supercooled 2D Yukawa liquids, the distribution function of dust particles displacement has the most discrepancy from the Gaussian distribution. In Section D, we discuss the origin of the non-Gaussian behavior.

At very long times limit, the $G_s$(r,$t$) curves are matched with Gaussian distributions, and relative differences between $G_s(r,t)$ and $G_s^{Gauss}(r,t)$ approach zero.\\

\begin{figure*}[t]
\begin{minipage}[b]{.32\textwidth}
\includegraphics[width=6.5cm,height=7.25cm]{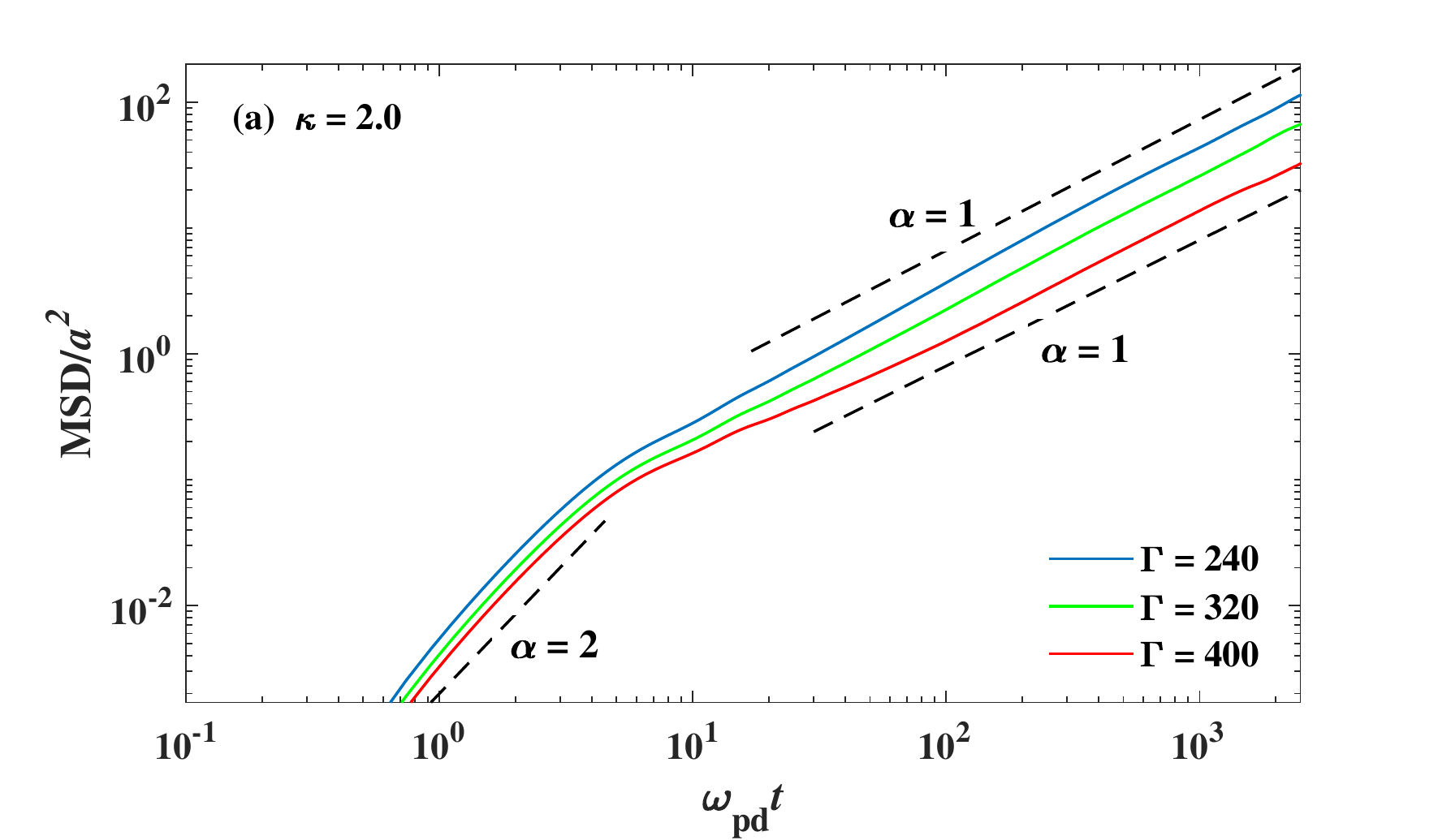}
\end{minipage}\qquad
\begin{minipage}[b]{0.34\textwidth}
\includegraphics[width=6.5cm,height=7.25cm]{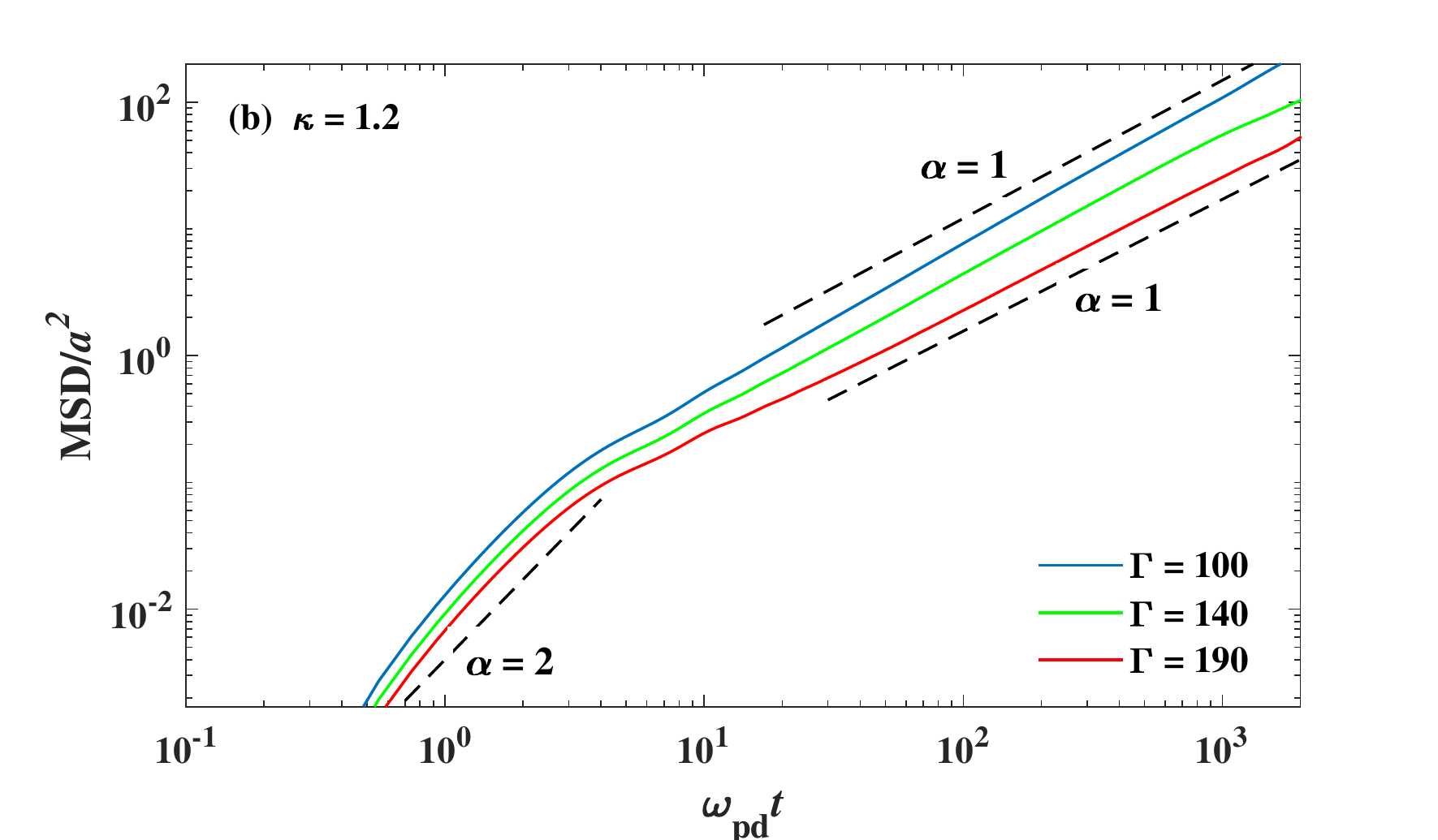}
\end{minipage}\qquad
\begin{minipage}[b]{.26\textwidth}
\includegraphics[width=6.2cm,height=7.25cm]{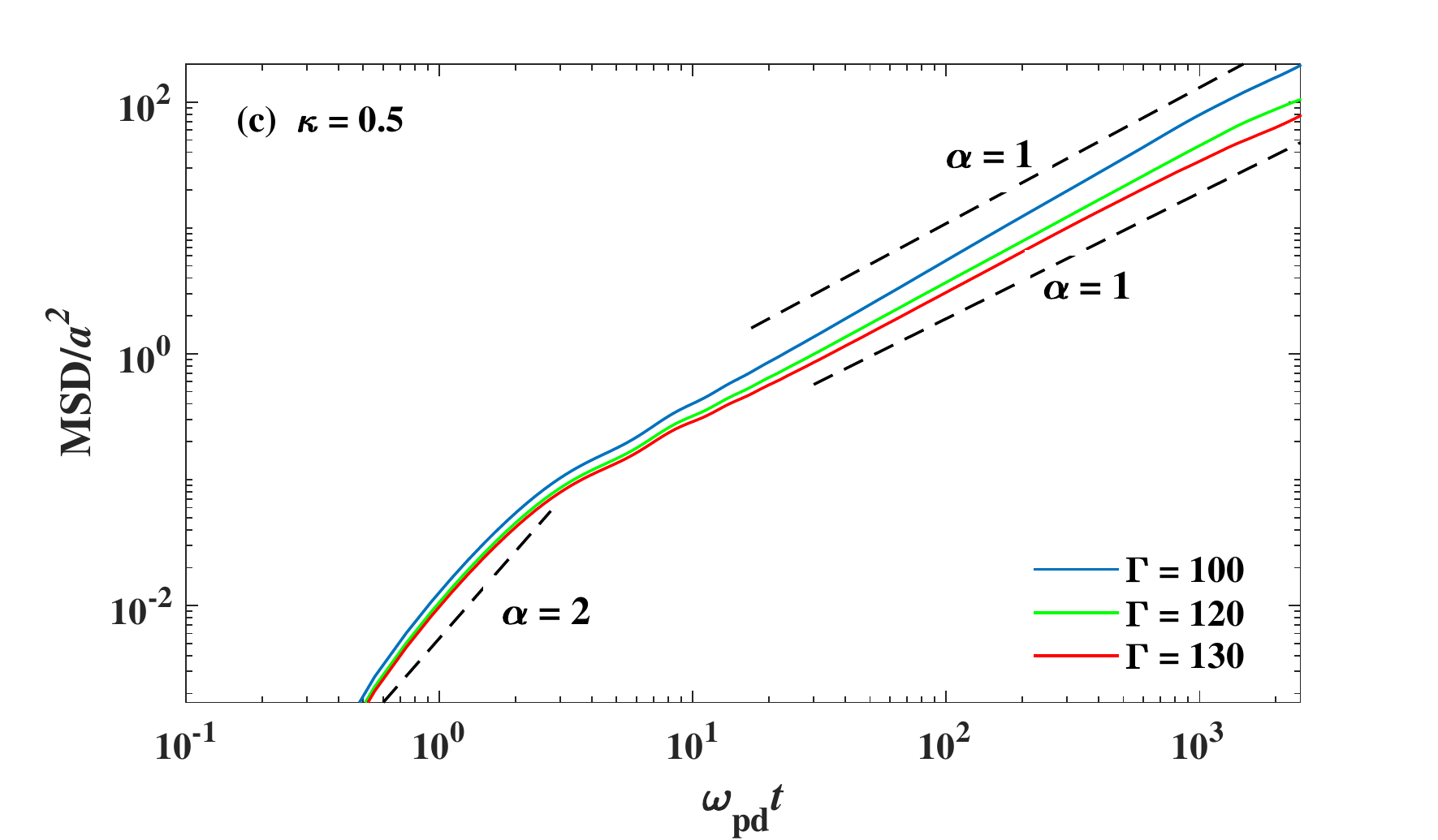}
\end{minipage}\qquad
\caption{\label{figure1} Mean-squared displacements $\langle(\Delta \mathrm{r}(t))^2\rangle $ for different temperatures. (a) $\kappa=2.0$, (b) $\kappa=1.2$, and (c) $\kappa=0.5$. At short times, MSD $\propto t^2$. At later times, when $100<\omega_{pd}t<600$, MSD $\propto t^{\alpha}$, where $\alpha$ comes very close to 1 with decreasing the temperature (increasing $\Gamma$), indicating Fickian diffusion.}

\end{figure*}

\begin{figure}[!htp]

\raggedleft
\includegraphics[width=9cm,height=7.25cm]{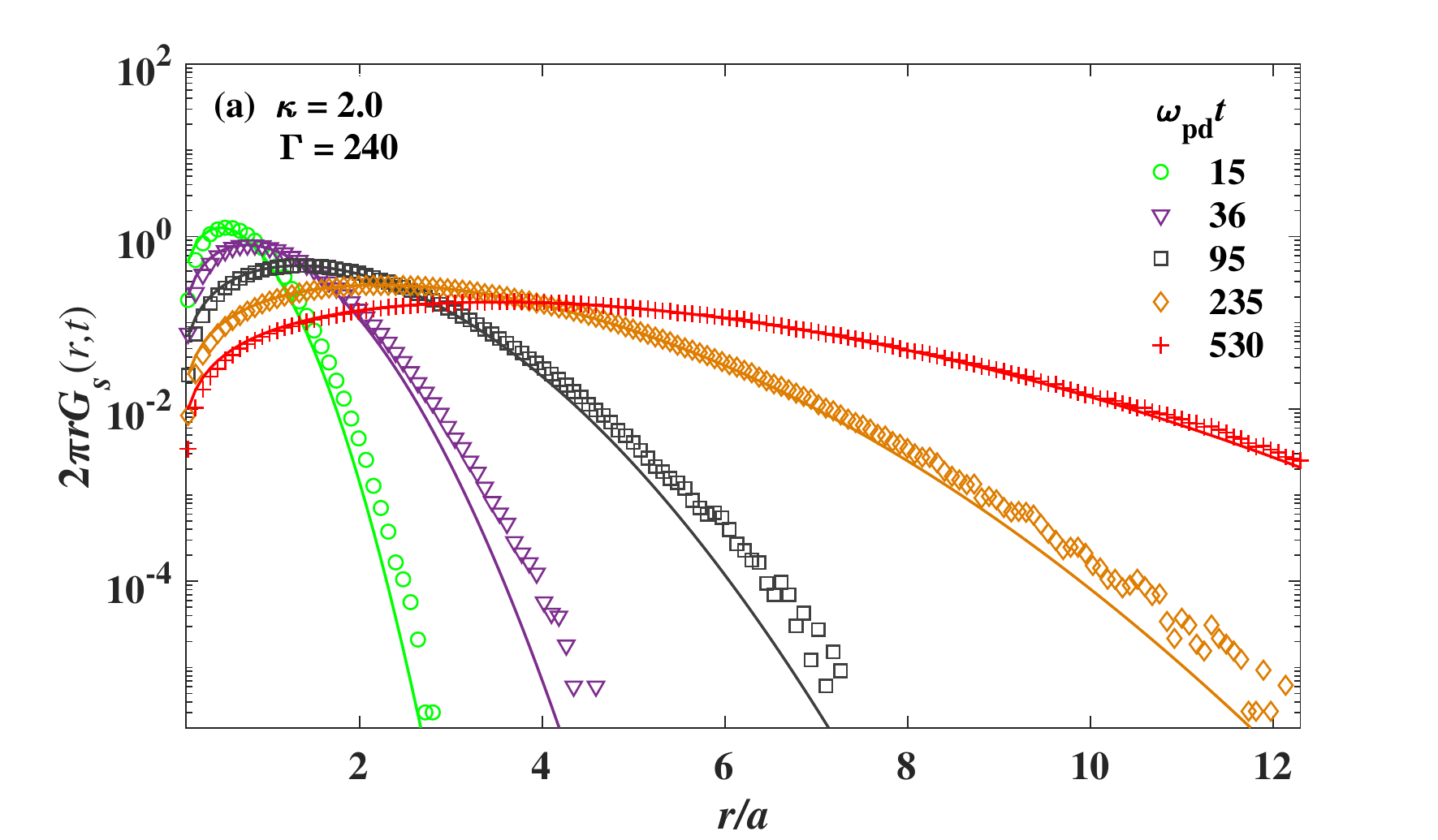}

\includegraphics[width=9cm,height=7.25cm]{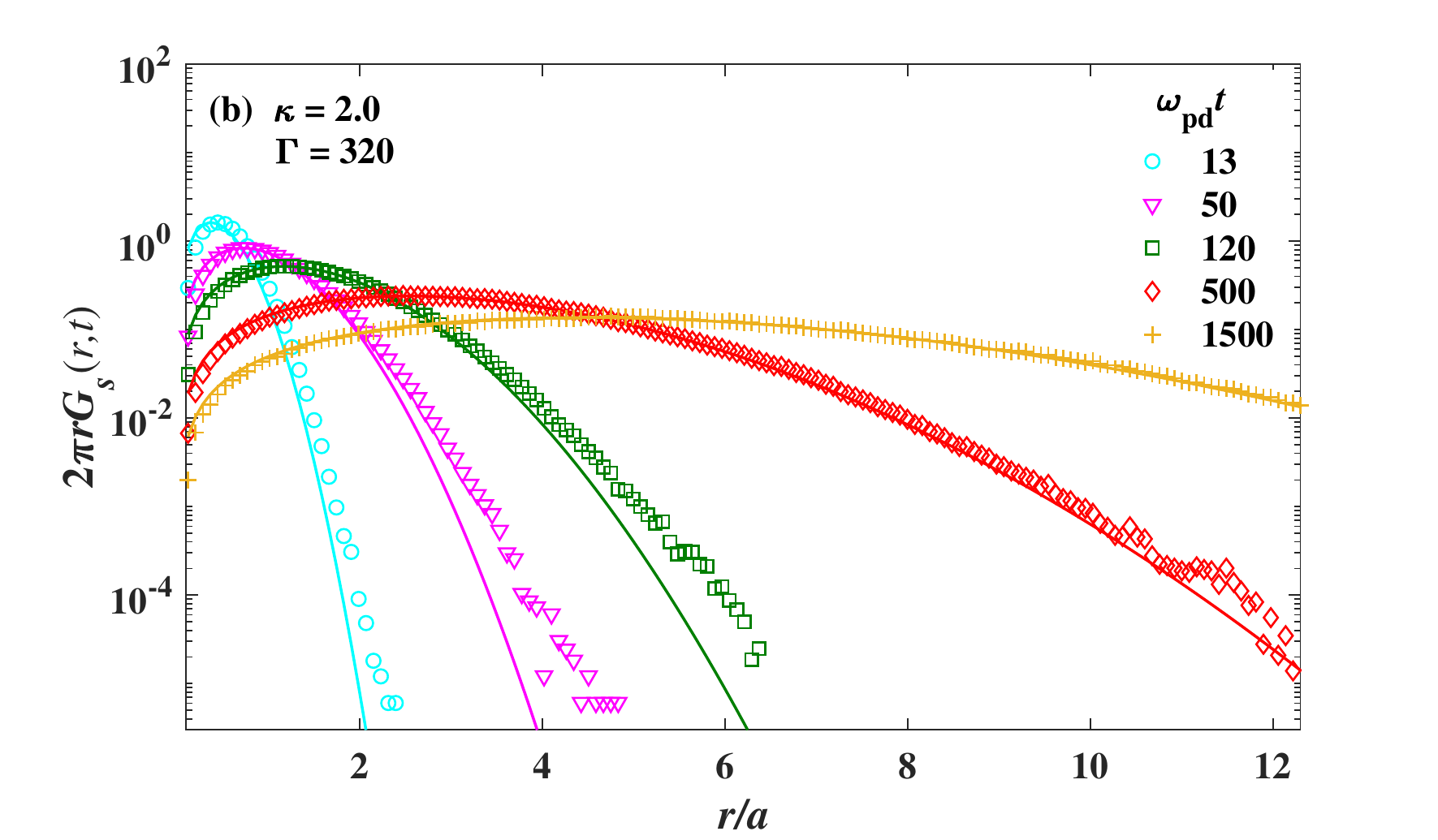}

\includegraphics[width=9cm,height=7.25cm]{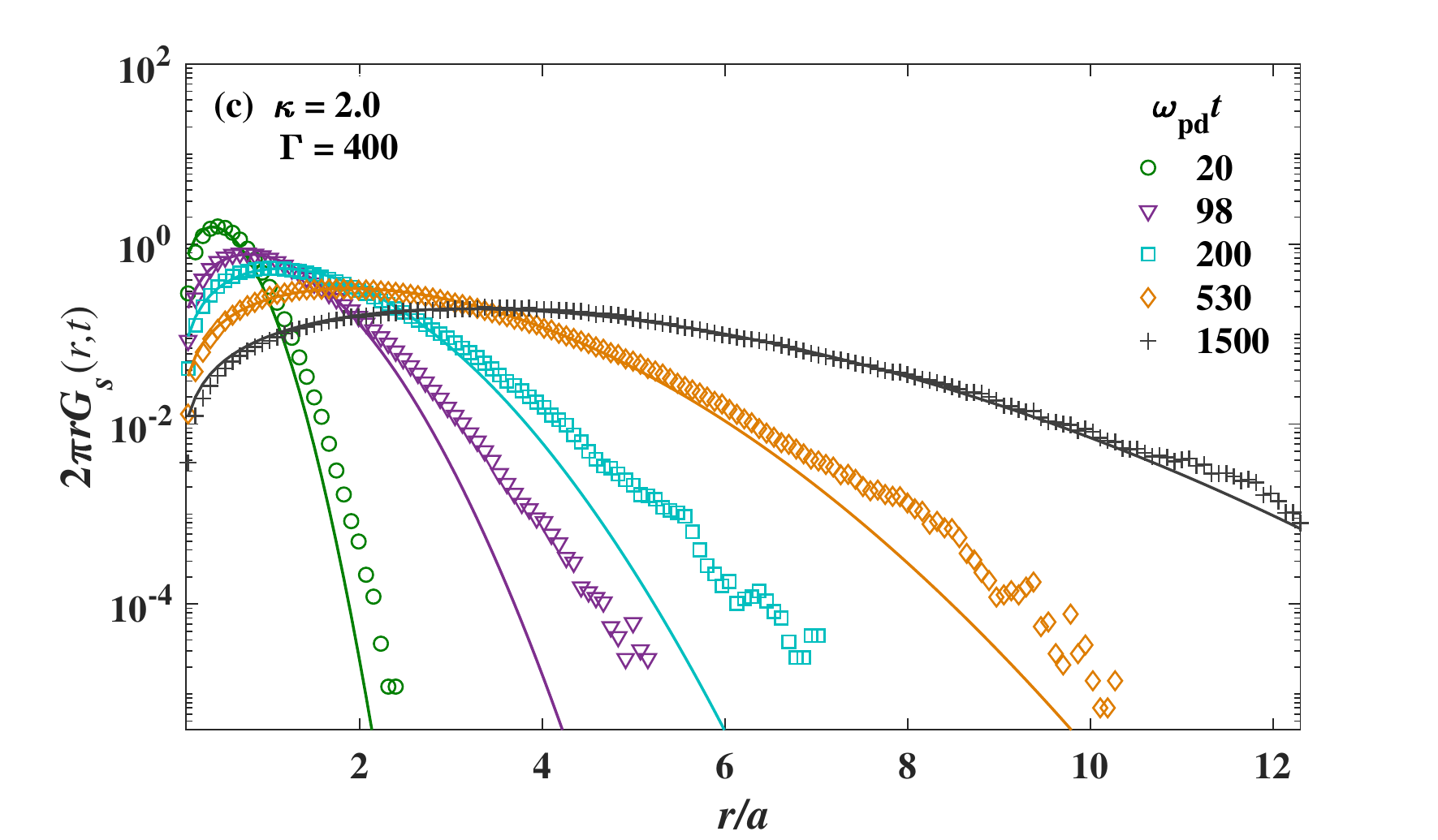}
\caption{Time evolution of the self-part of van Hove functions for $\kappa=2.0$. (a) $\Gamma=240$, (b) $\Gamma=320$, (c) $\Gamma=400$. The symbols are the simulation results, and solid lines are from the Gaussian distribution in Eq. (\ref{eq2}) with the MSD obtained from the simulation.}\label{label2}
\end{figure}

\begin{figure}[!htp]

\includegraphics[width=9cm,height=7.25cm]{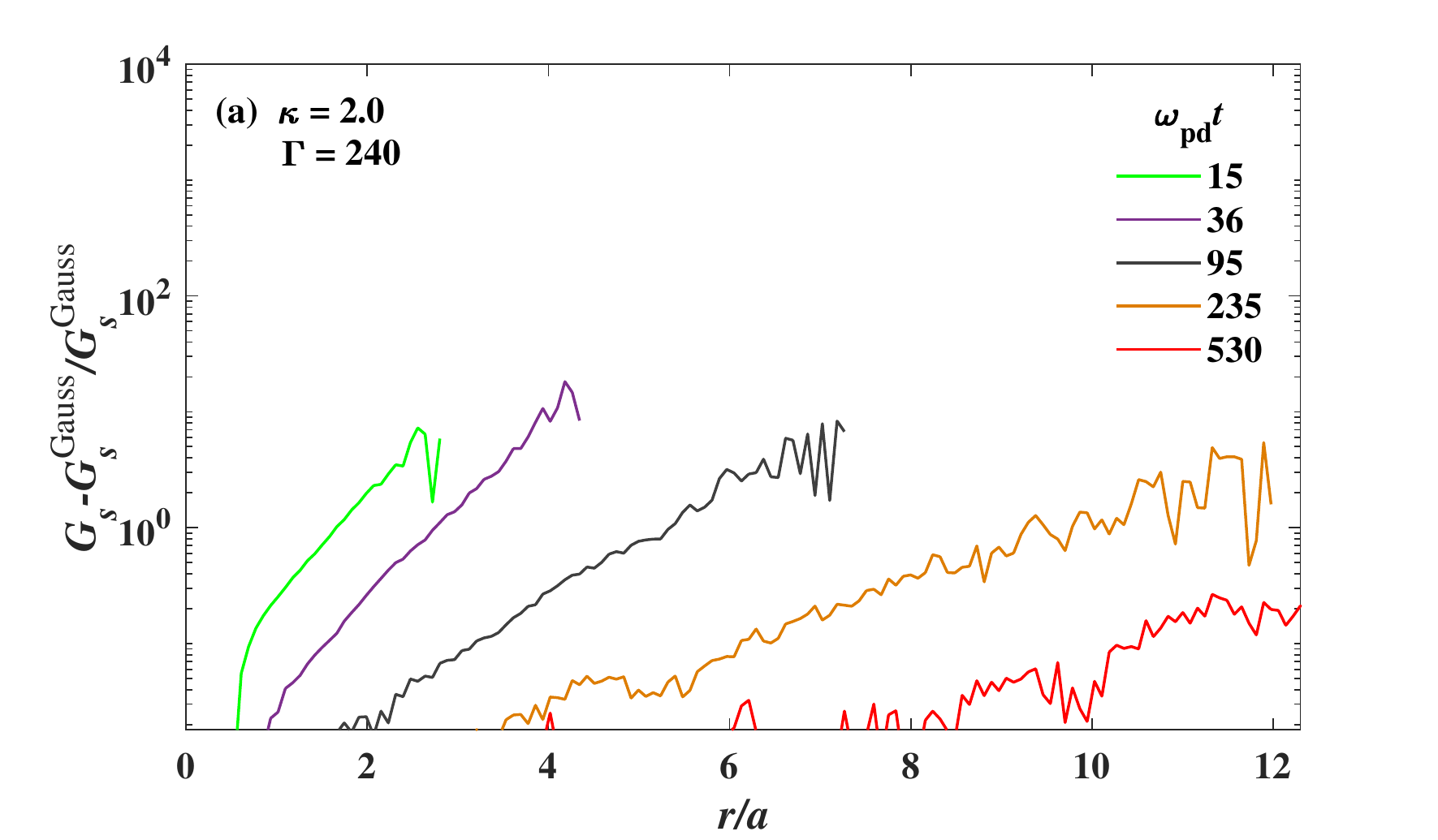}

\includegraphics[width=9cm,height=7.25cm]{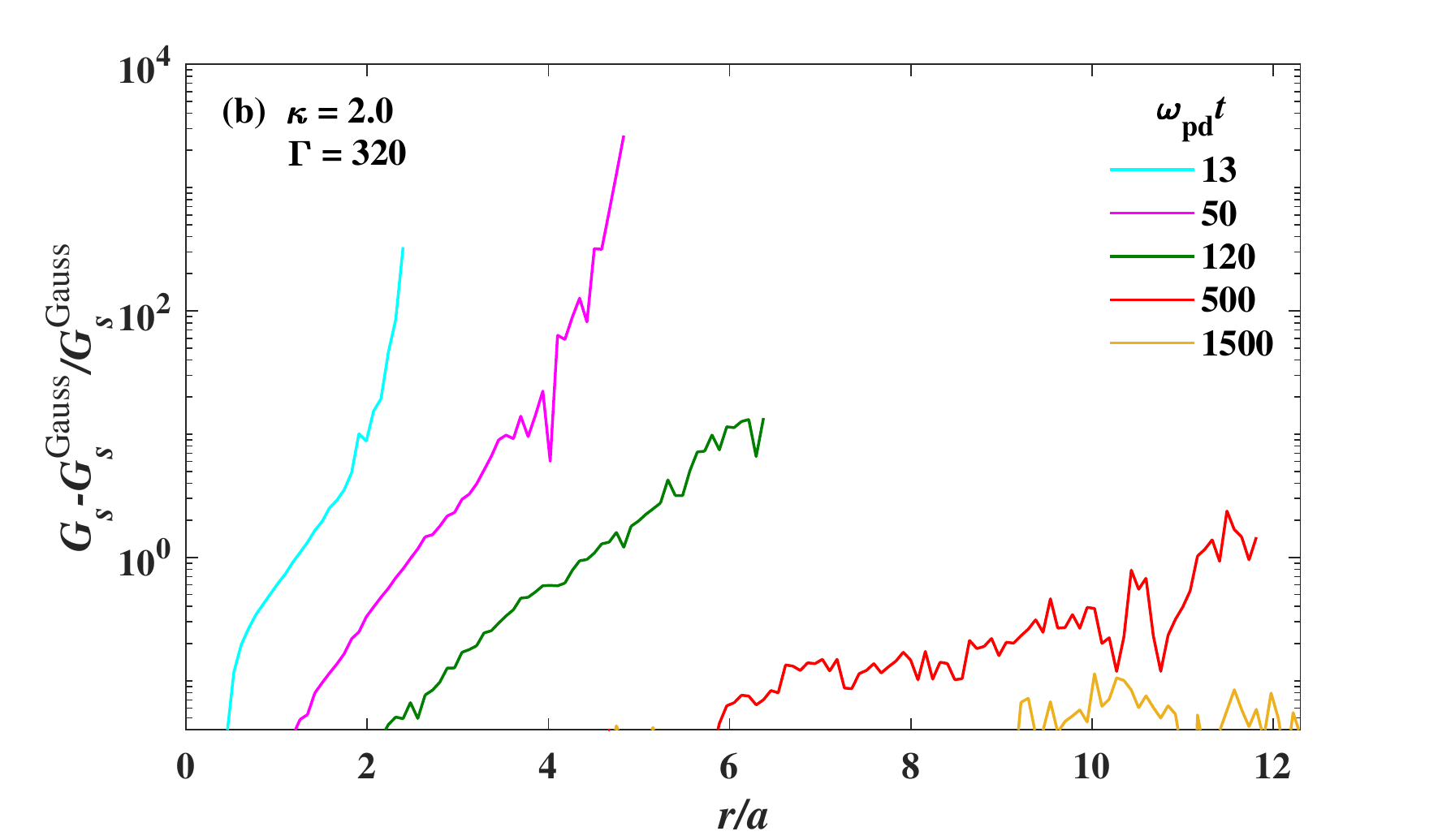}

\includegraphics[width=9cm,height=7.25cm]{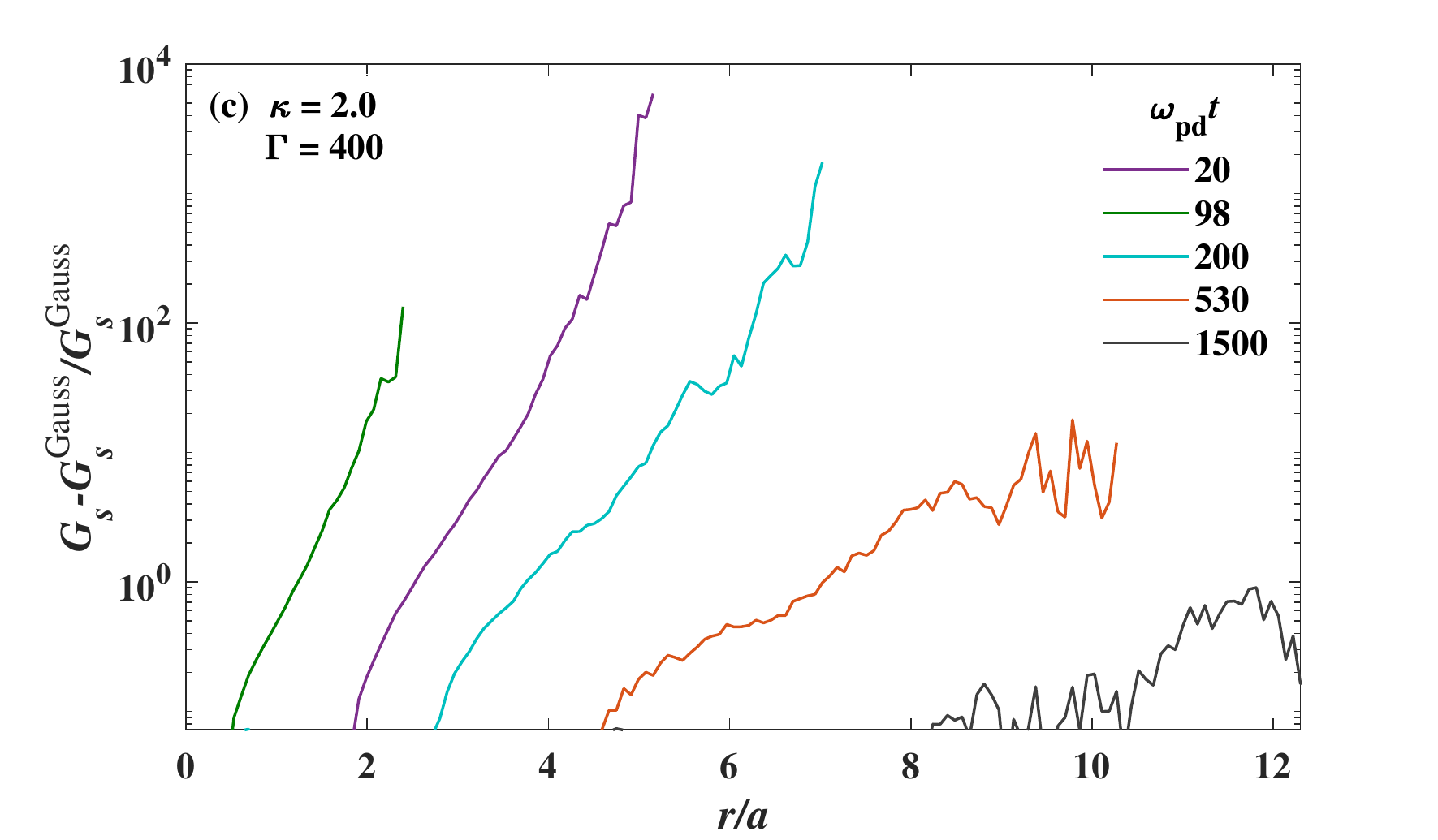}

\caption{Relative differences between $G_s\rm{(}\it{r},t\rm{)}$ and Gaussian distribution $G_s^{\rm{Gauss}}\rm{(}\it{r},t\rm{)}$ (obtained from Eq.~(\ref{eq2})) for (a) $\Gamma=240$, (b) $\Gamma=320$, (c) $\Gamma=400$.
}\label{label3}
\end{figure}
\begin{figure}[!htp]

\includegraphics[width=9cm,height=7.25cm]{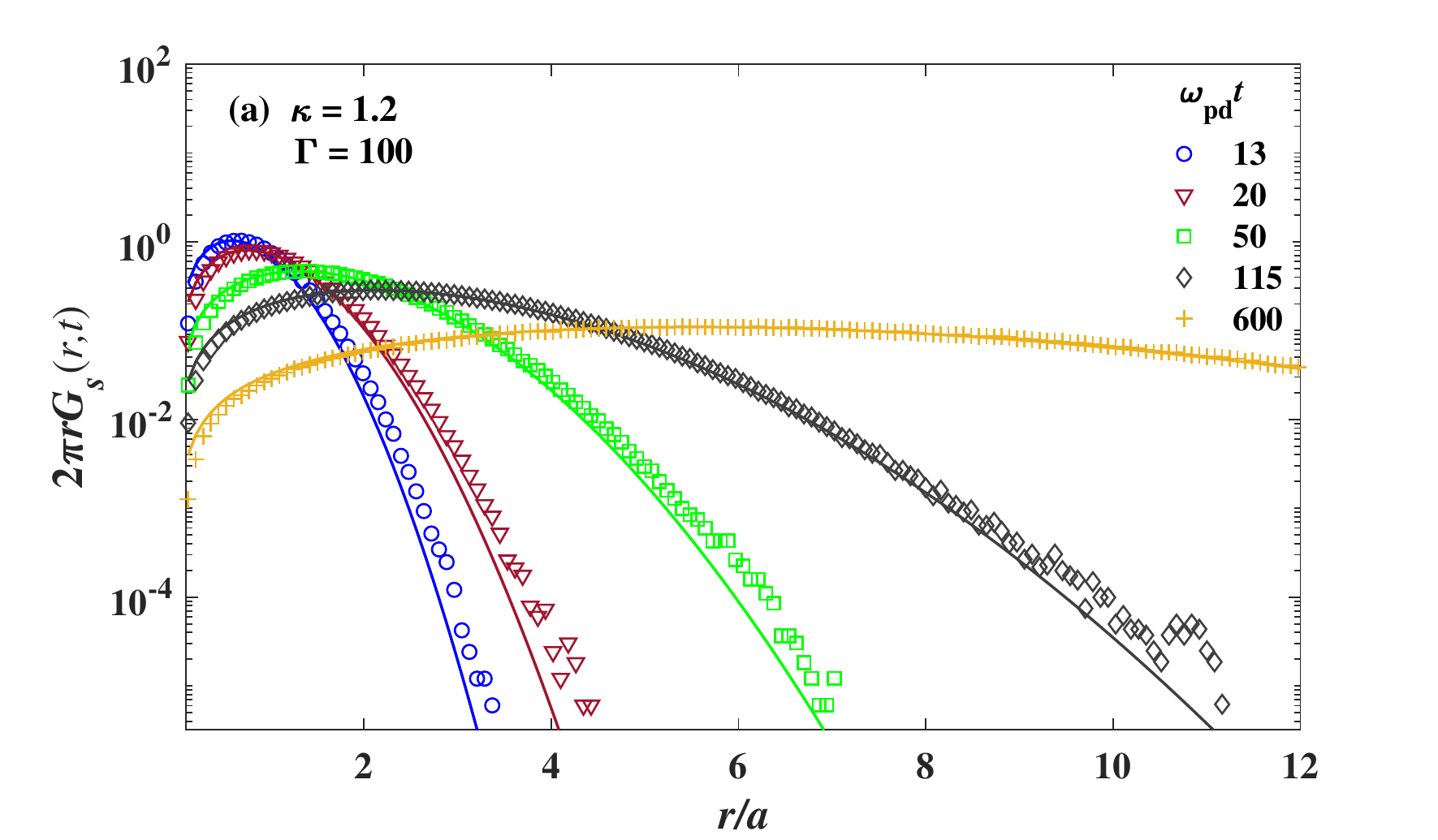}

\includegraphics[width=9cm,height=7.25cm]{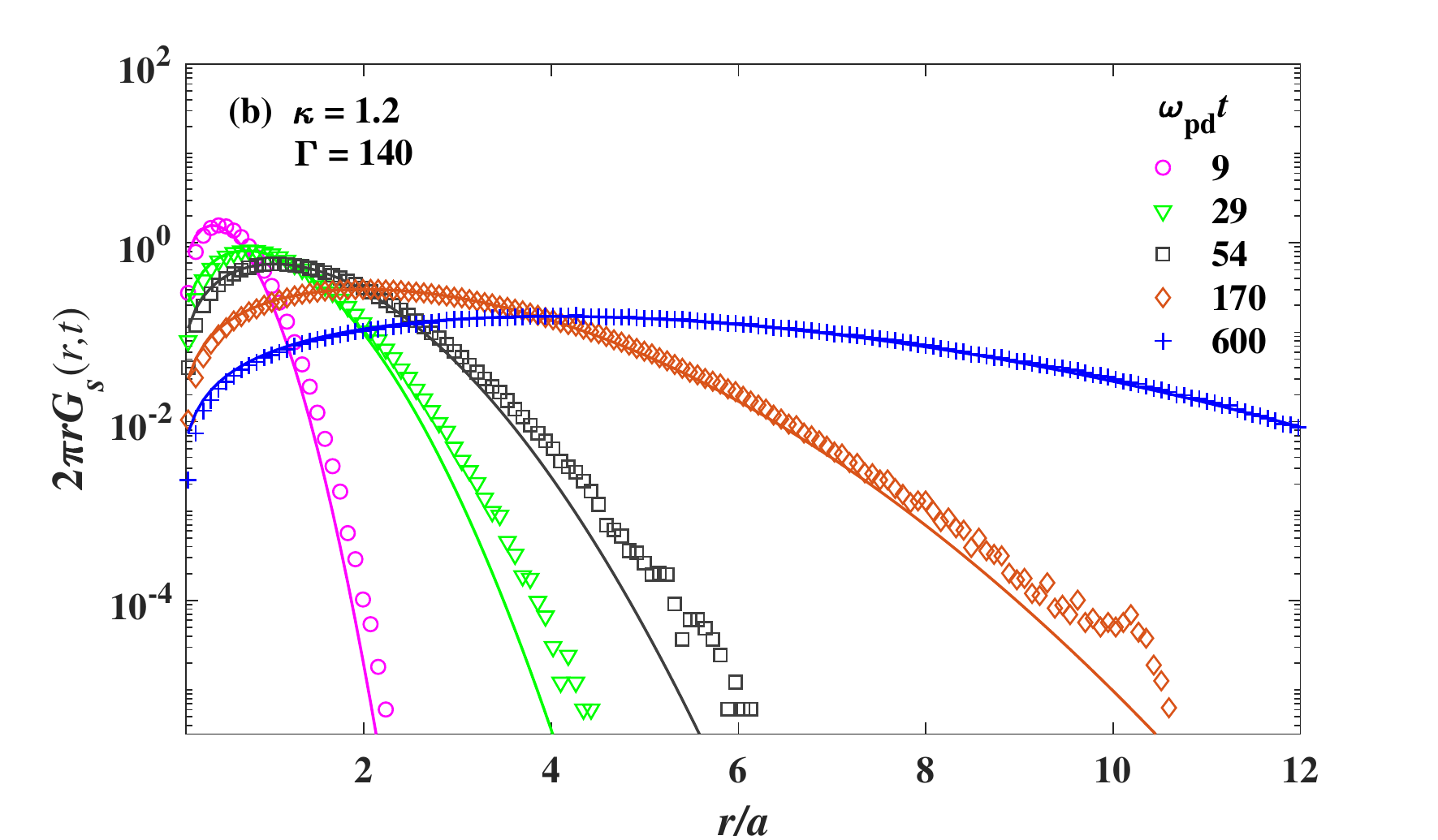}

\includegraphics[width=9cm,height=7.25cm]{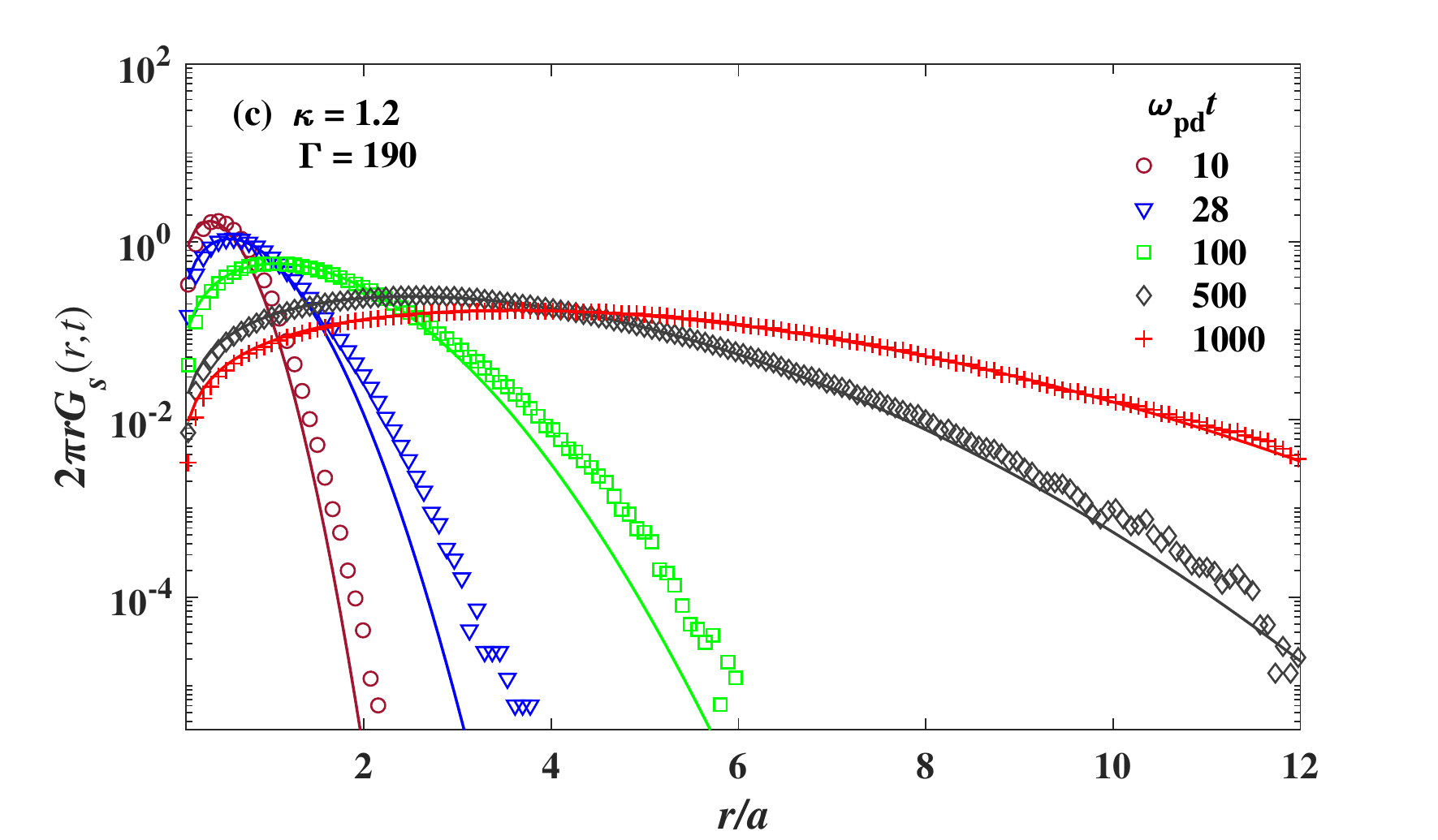}

\caption{Time evolution of the self-part of van Hove functions for $\kappa=1.2$. (a) $\Gamma=100$, (b) $\Gamma=140$, (c) $\Gamma=190$. The symbols are the simulation results, and solid lines are from the Gaussian distribution in Eq. (\ref{eq2}) with the MSD obtained from the simulation.}\label{label4}

\end{figure}
\begin{figure}[!htp]

\includegraphics[width=9cm,height=7.25cm]{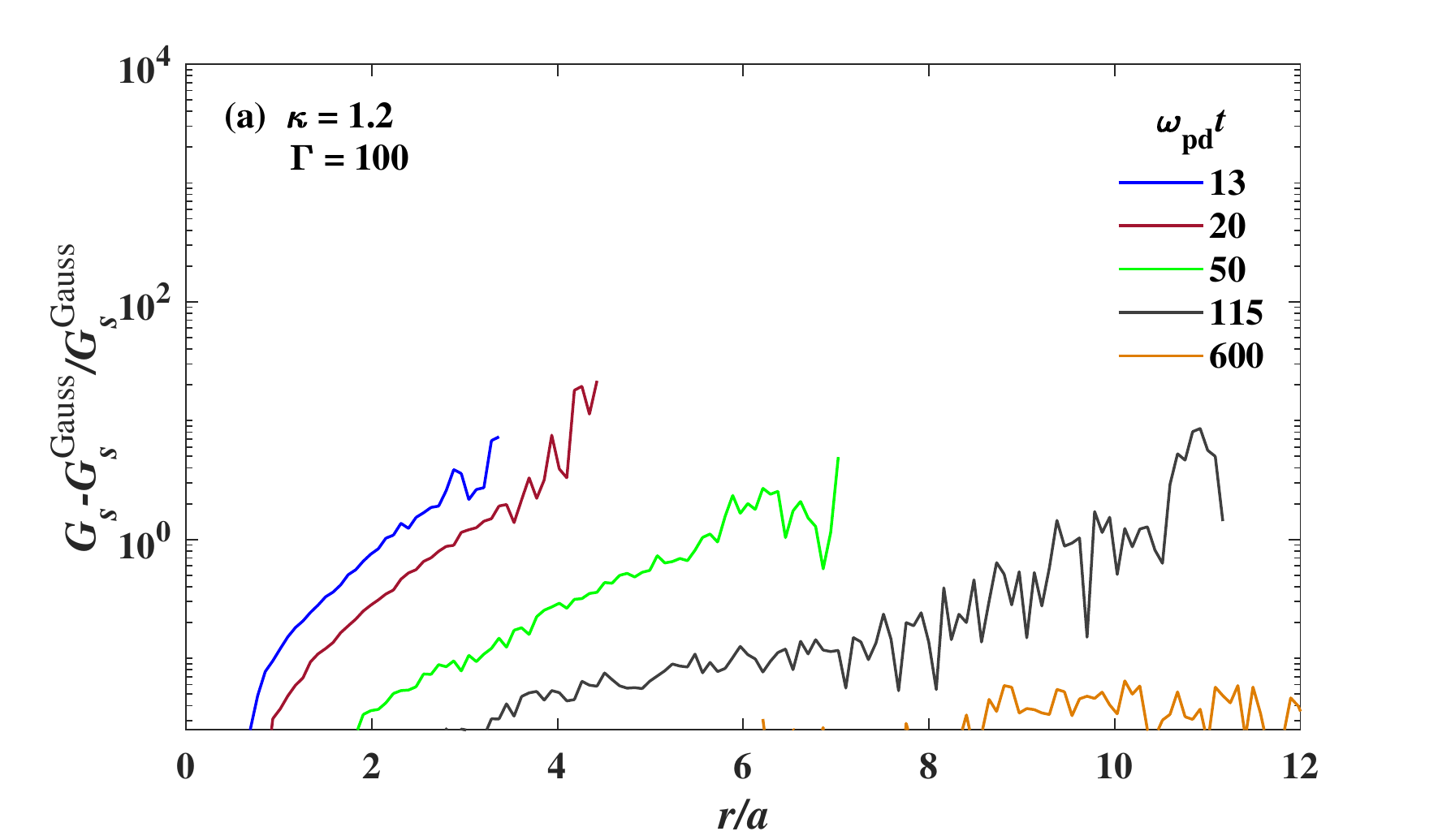}

\includegraphics[width=9cm,height=7.25cm]{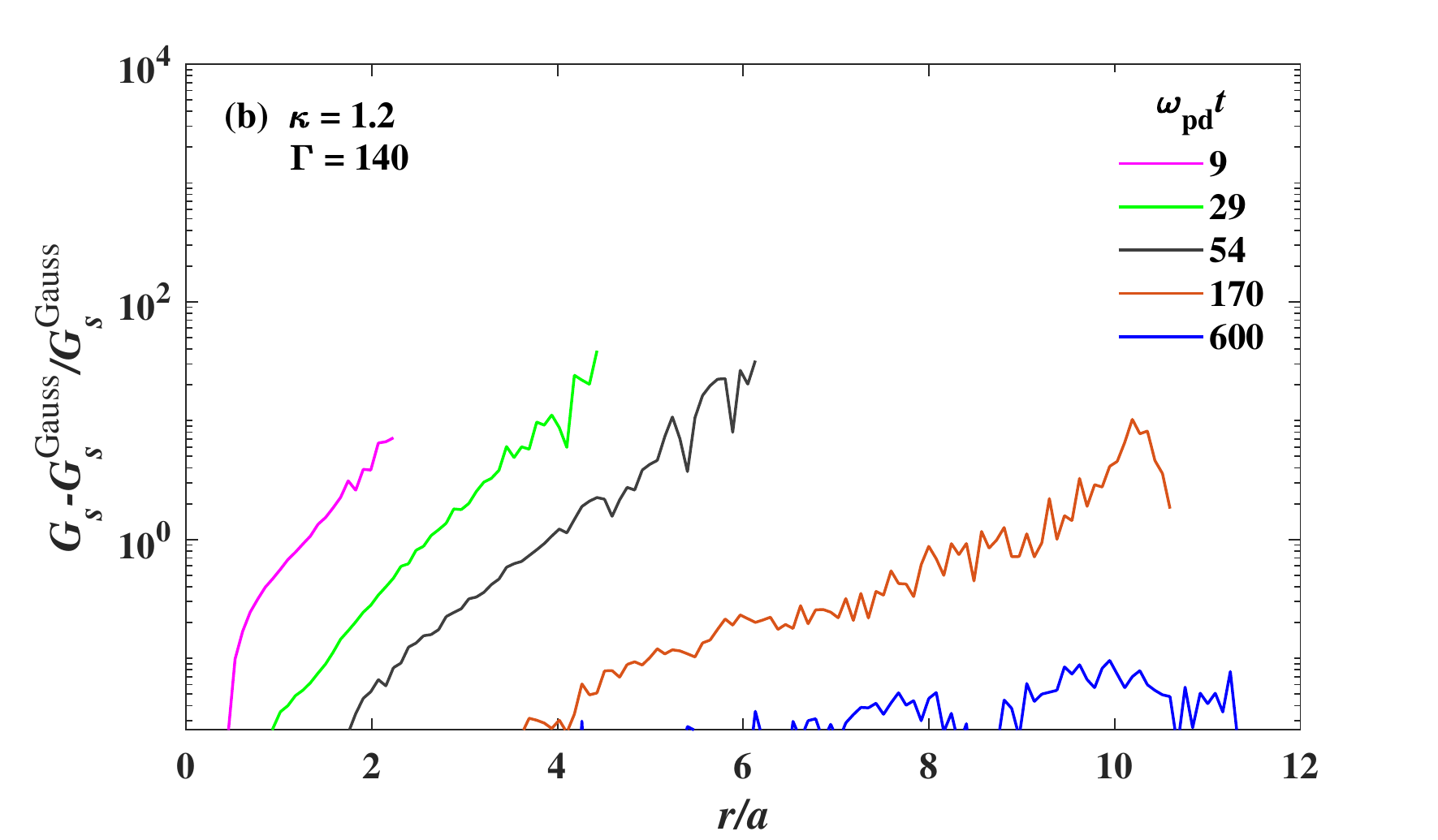}

\includegraphics[width=9cm,height=7.25cm]{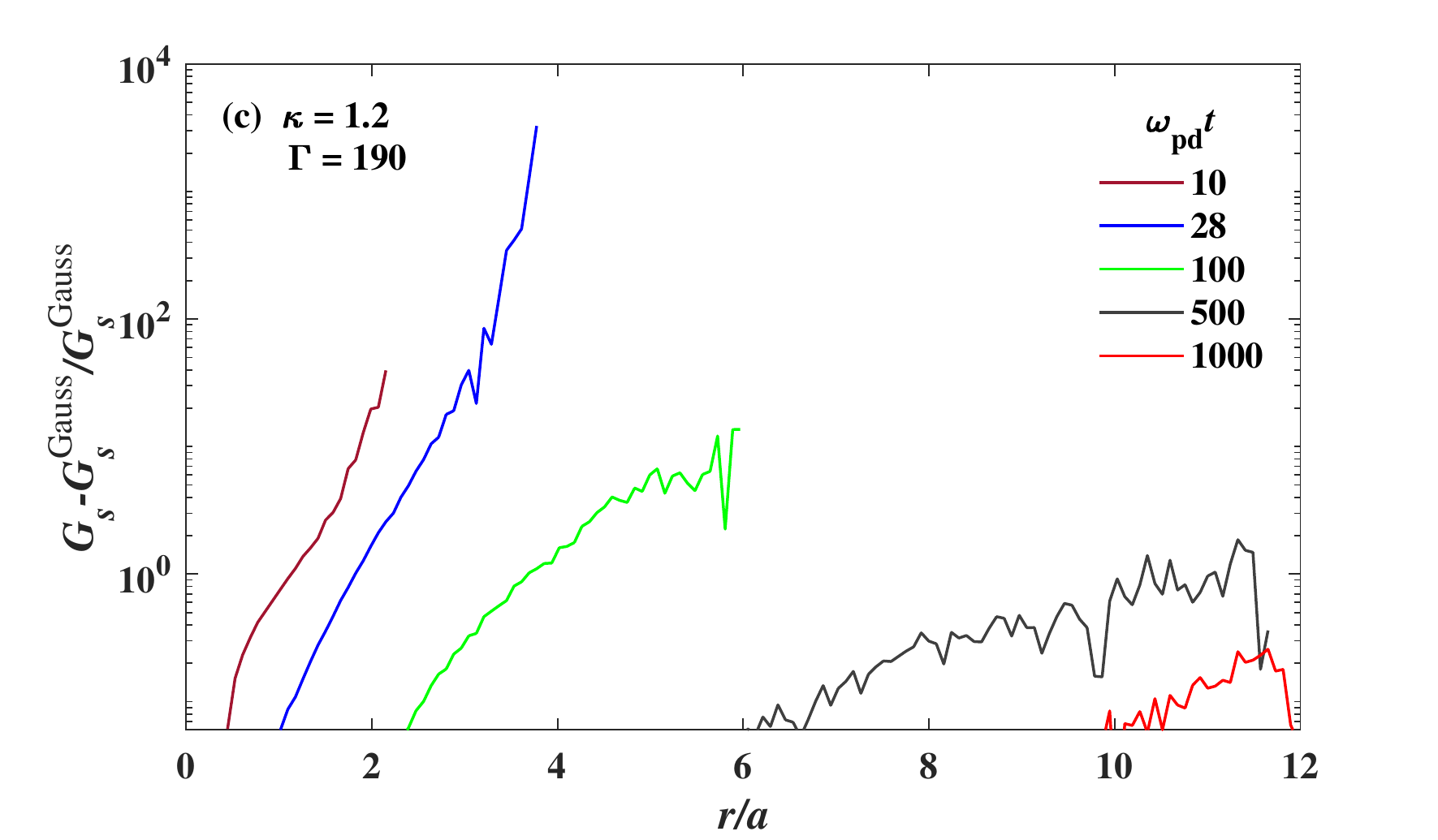}

\caption{Relative differences between $G_s\rm{(}\it{r},t\rm{)}$ and Gaussian distribution $G_s^{\rm{Gauss}}\rm{(}\it{r},t\rm{)}$ (obtained from Eq.~(\ref{eq2})) for (a) $\Gamma=100$, (b) $\Gamma=140$, (c) $\Gamma=190$.
}\label{label5}
\end{figure}

\begin{figure}[!htp]

\includegraphics[width=9cm,height=7.25cm]{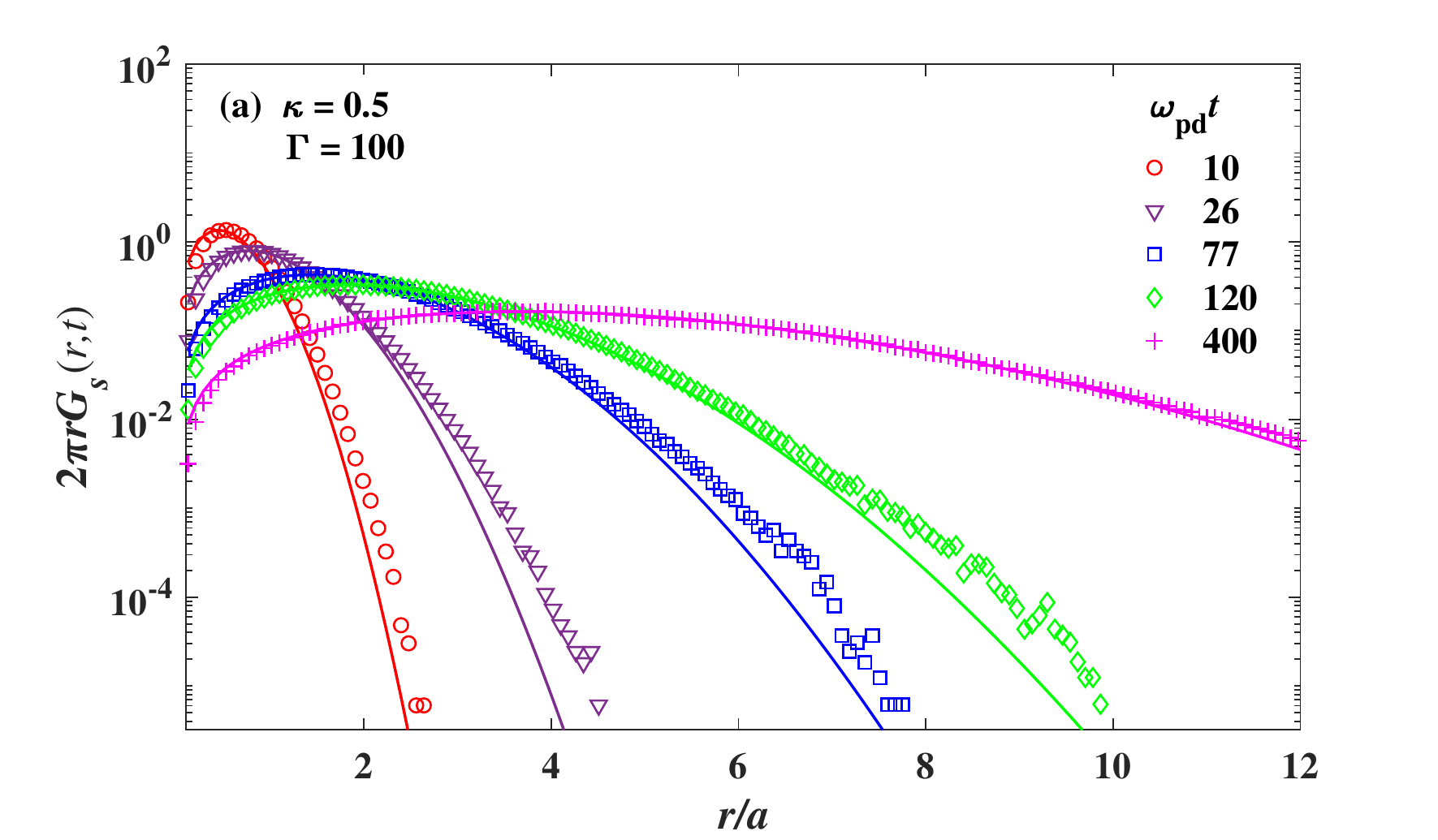}

\includegraphics[width=9cm,height=7.25cm]{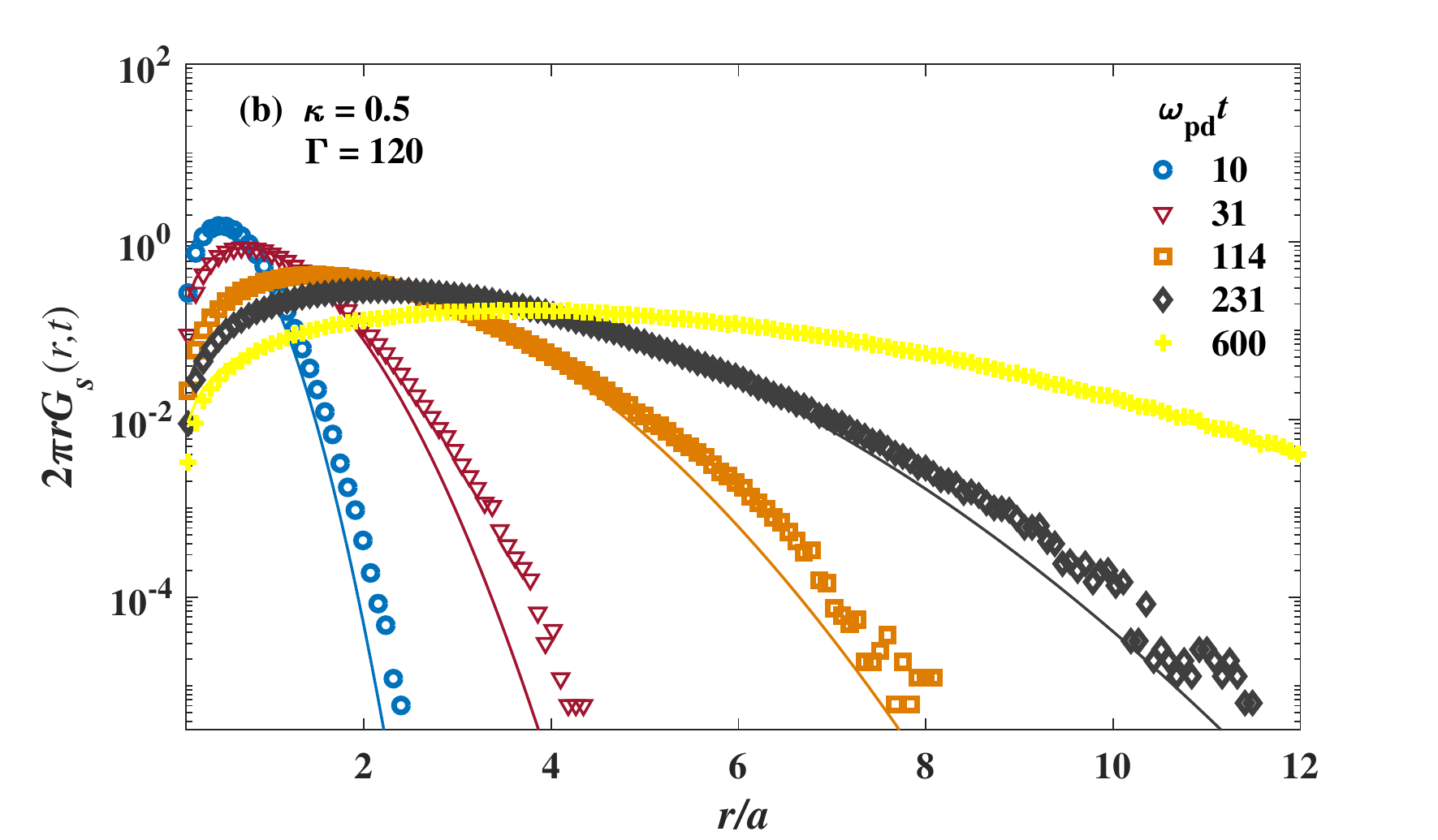}

\includegraphics[width=9cm,height=7.25cm]{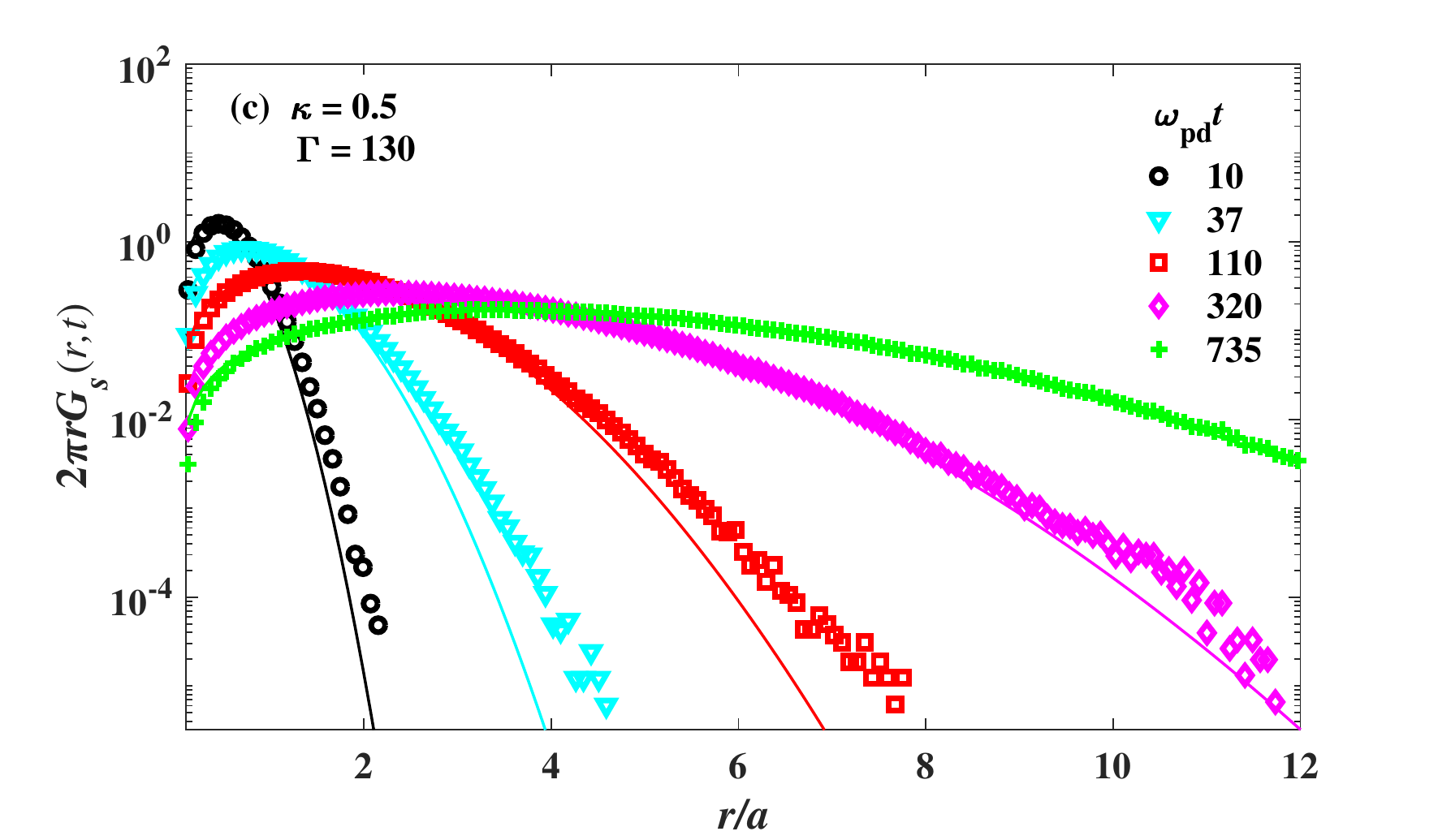}
\caption{Time evolution of the self-part of van Hove functions for $\kappa=0.5$. (a) $\Gamma=100$, (b) $\Gamma=120$, (c) $\Gamma=130$. The symbols are the simulation results, and solid lines are from the Gaussian distribution in Eq. (\ref{eq2}) with the MSD obtained from the simulation.
}\label{label6}

\end{figure}
\begin{figure}[!htp]

\includegraphics[width=9cm,height=7.25cm]{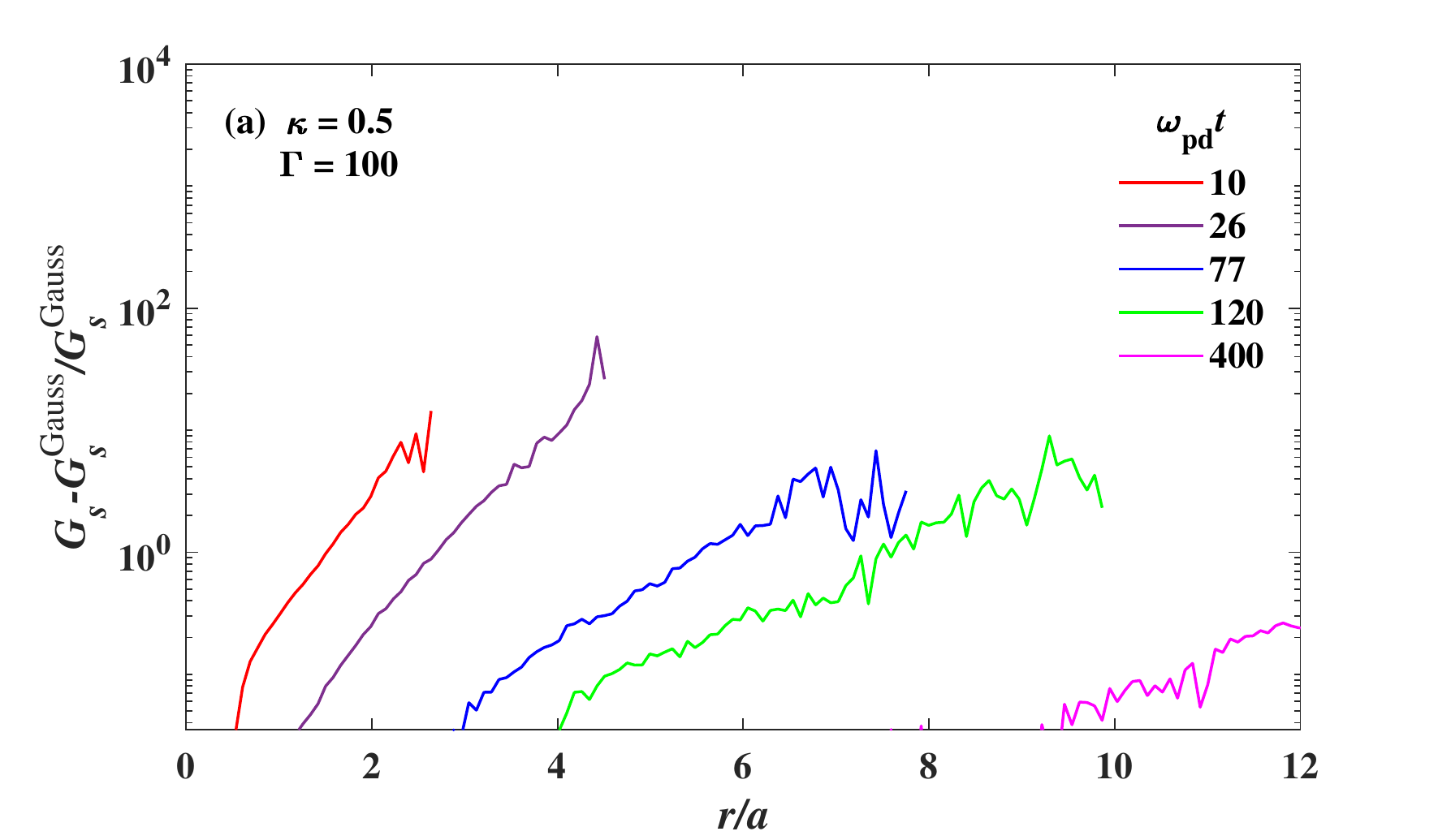}

\includegraphics[width=9cm,height=7.25cm]{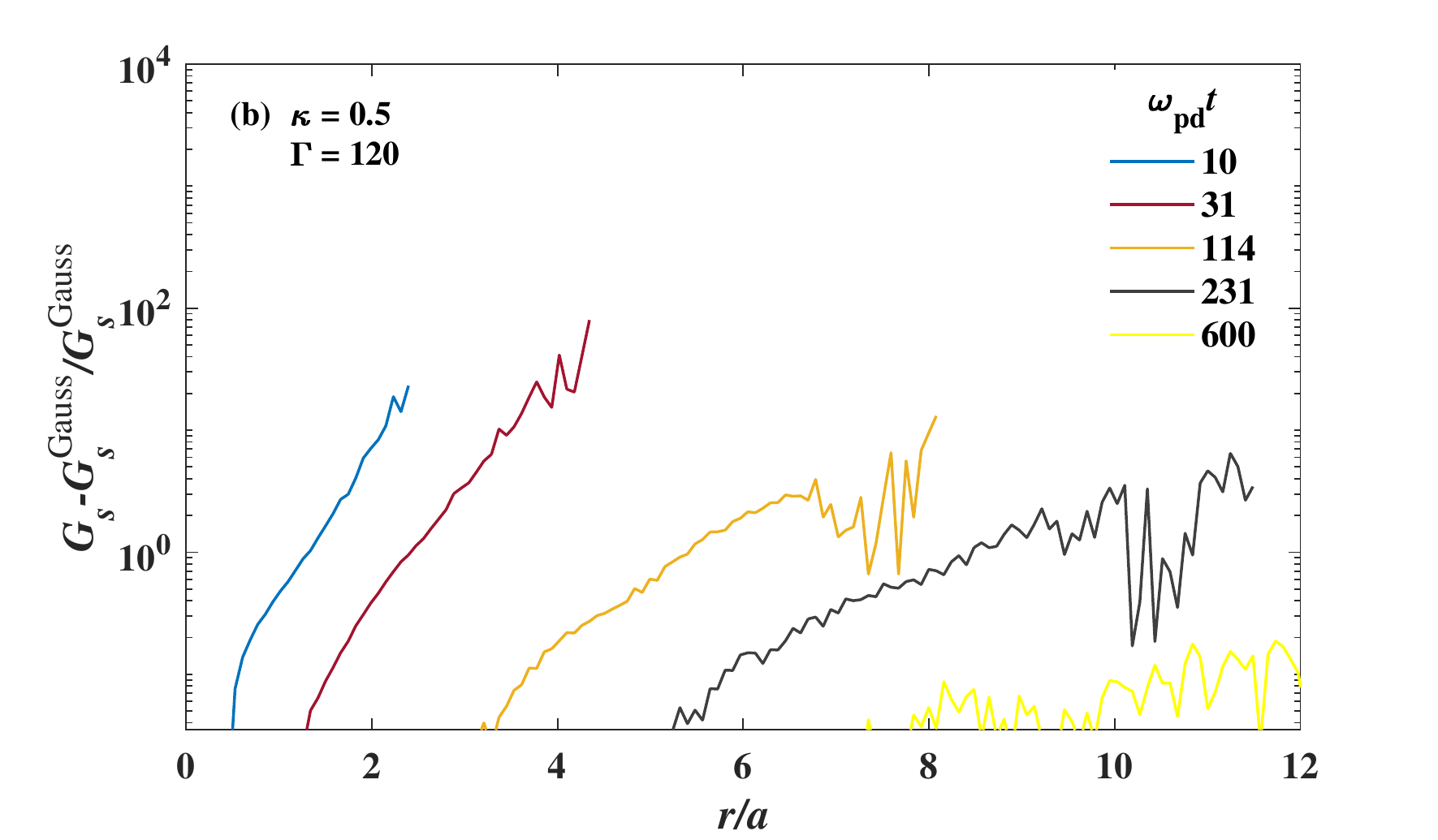}

\includegraphics[width=9cm,height=7.25cm]{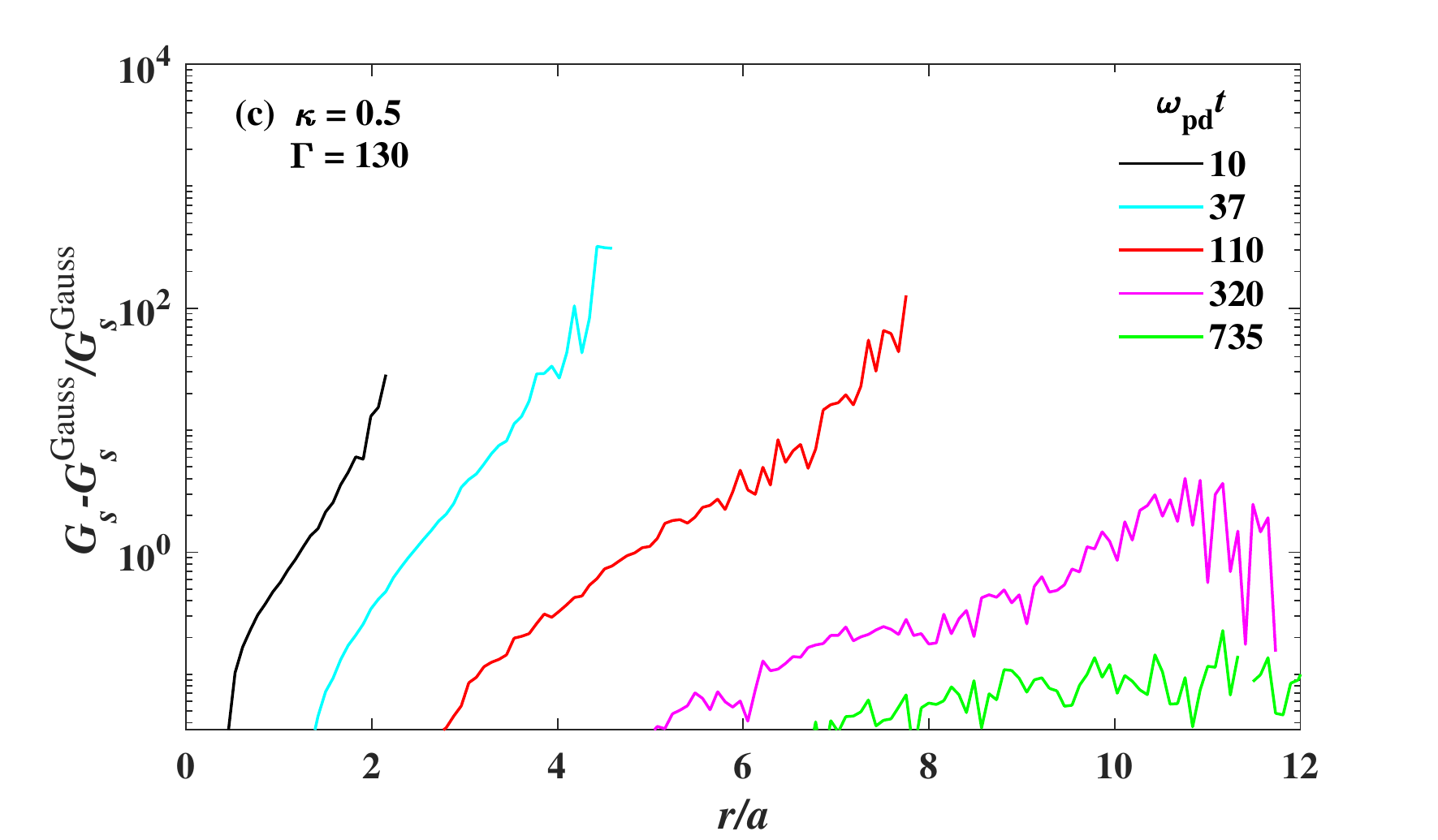}

\caption{Relative differences between $G_s\rm{(}\it{r},t\rm{)}$ and Gaussian distribution $G_s^{\rm{Gauss}}\rm{(}\it{r},t\rm{)}$ (obtained from Eq.~(\ref{eq2})) for (a)$\Gamma=100$, (b) $\Gamma=120$, (c) $\Gamma=130$.
}\label{label7}
\end{figure}

\subsection{\label{NGP}Non-Gaussian parameter}
Deviations from a Gaussian are quantified by a non-Gaussian parameter (NGP). For a 2D system, the NGP is given by~\cite{Rahman1964}
\begin{equation}
\alpha_2(t)=\frac{1}{2}\frac{\langle(\Delta \mathrm{r}(t))^4\rangle}{[\langle(\Delta \mathrm{r}(t))^2\rangle]^2}-1,
\label{eq7}
\end{equation} 
where the moments $\langle(\Delta \mathrm{r}(t))^n\rangle$ are defined as
\begin{equation}
\langle(\Delta \mathrm{r}(t))^n\rangle=\langle\vert\boldsymbol{\mathrm{r}}(t)-\boldsymbol{\mathrm{r}}(0)\vert^n\rangle=\int \mathrm{r}^n G_s(\boldsymbol{\mathrm{r}},t)d\boldsymbol{\mathrm{r}}\label{eq8}.
\end{equation}
Using these two equations, we can see that the NGP is exactly zero for a Gaussian distribution (given by Eq. (\ref{eq2})).

 Figures 8(a)--8(c) show the non-Gaussian parameter $\alpha_2(t)$ for the selected $\kappa$ and $\Gamma$ values.
The time evolution of the non-Gaussian parameter can be classified into three regimes:

(i) At short times when dust particles move in the cage created by neighboring dust particles, $\alpha_2(t)$ is zero.

(ii) At later times when dust particles escape from the cages and diffuse, $\alpha_2(t)$ increases with time so that it reaches a peak at intermediate times indicating maximum deviation from the Gaussian. For each $\kappa$ value, this peak increases with increasing $\Gamma$ equivalently decreasing $T$.

(iii) At very long times limit, $\alpha_2(t)$ decays to zero, as we expected from the Gaussian behavior of $G_s\rm{(}\it{r},t\rm{)}$ at these times.

\begin{figure}[!htp]

\includegraphics[width=9cm,height=7.25cm]{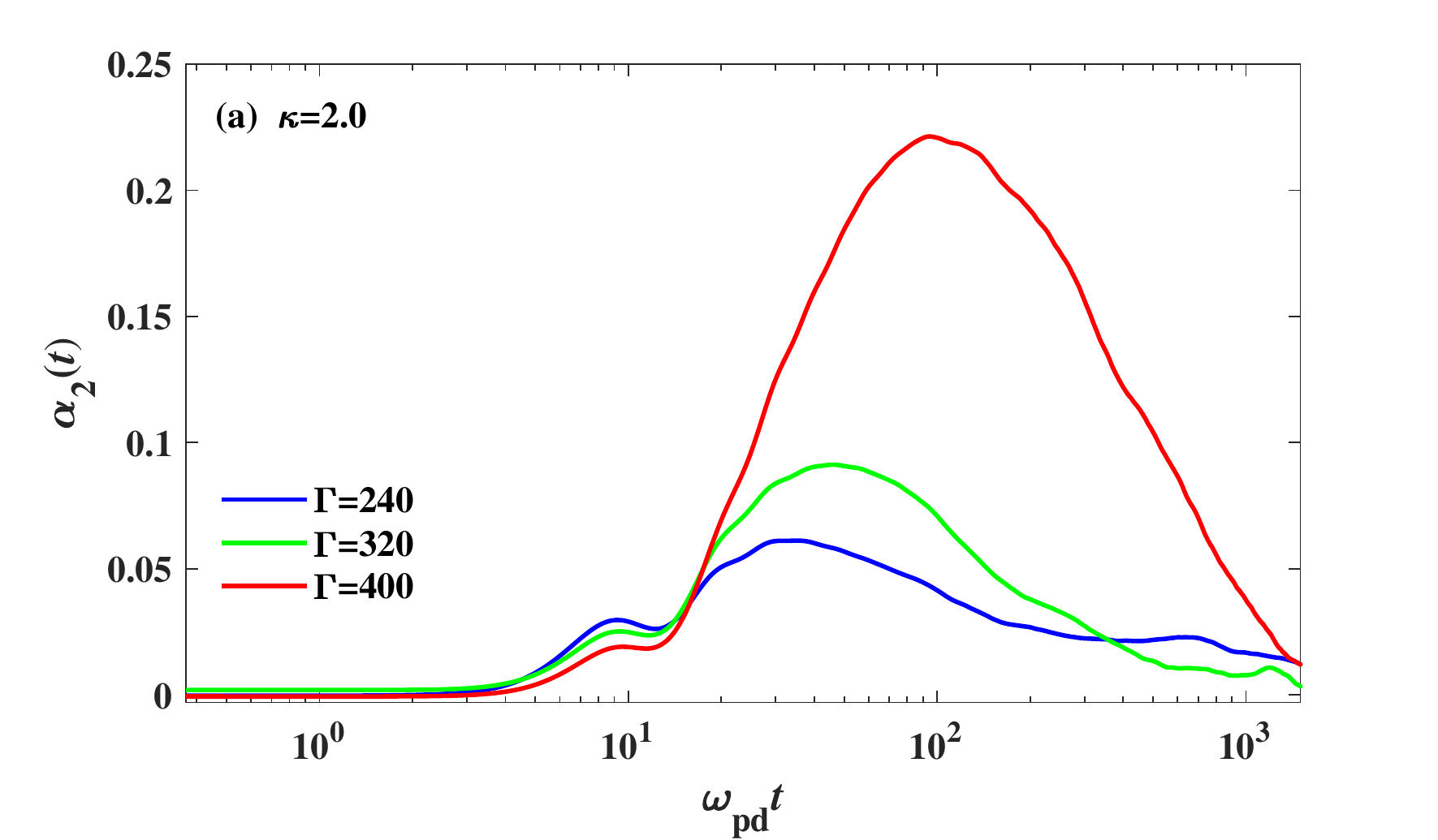}

\includegraphics[width=9cm,height=7.25cm]{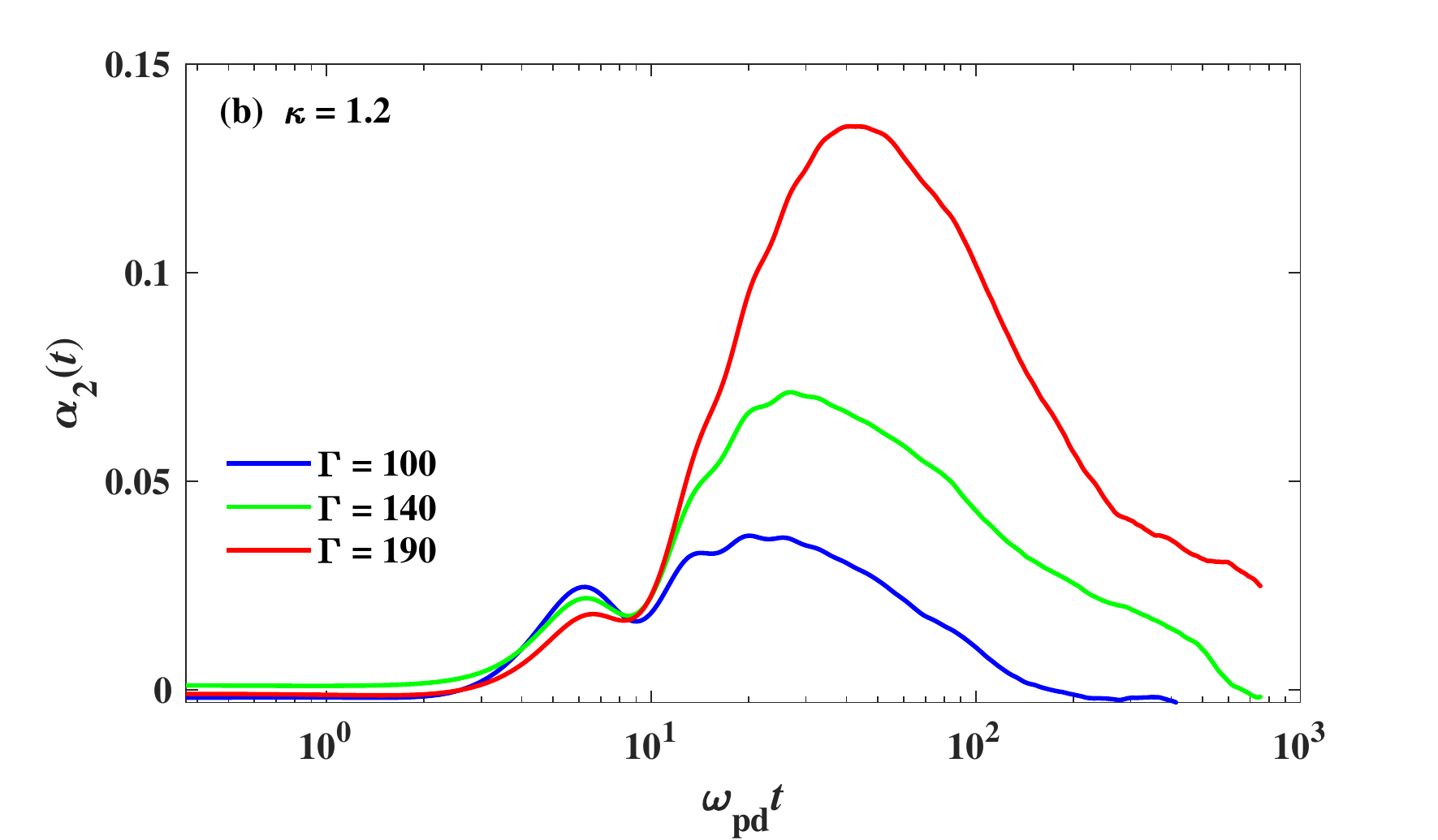}

\includegraphics[width=9cm,height=7.25cm]{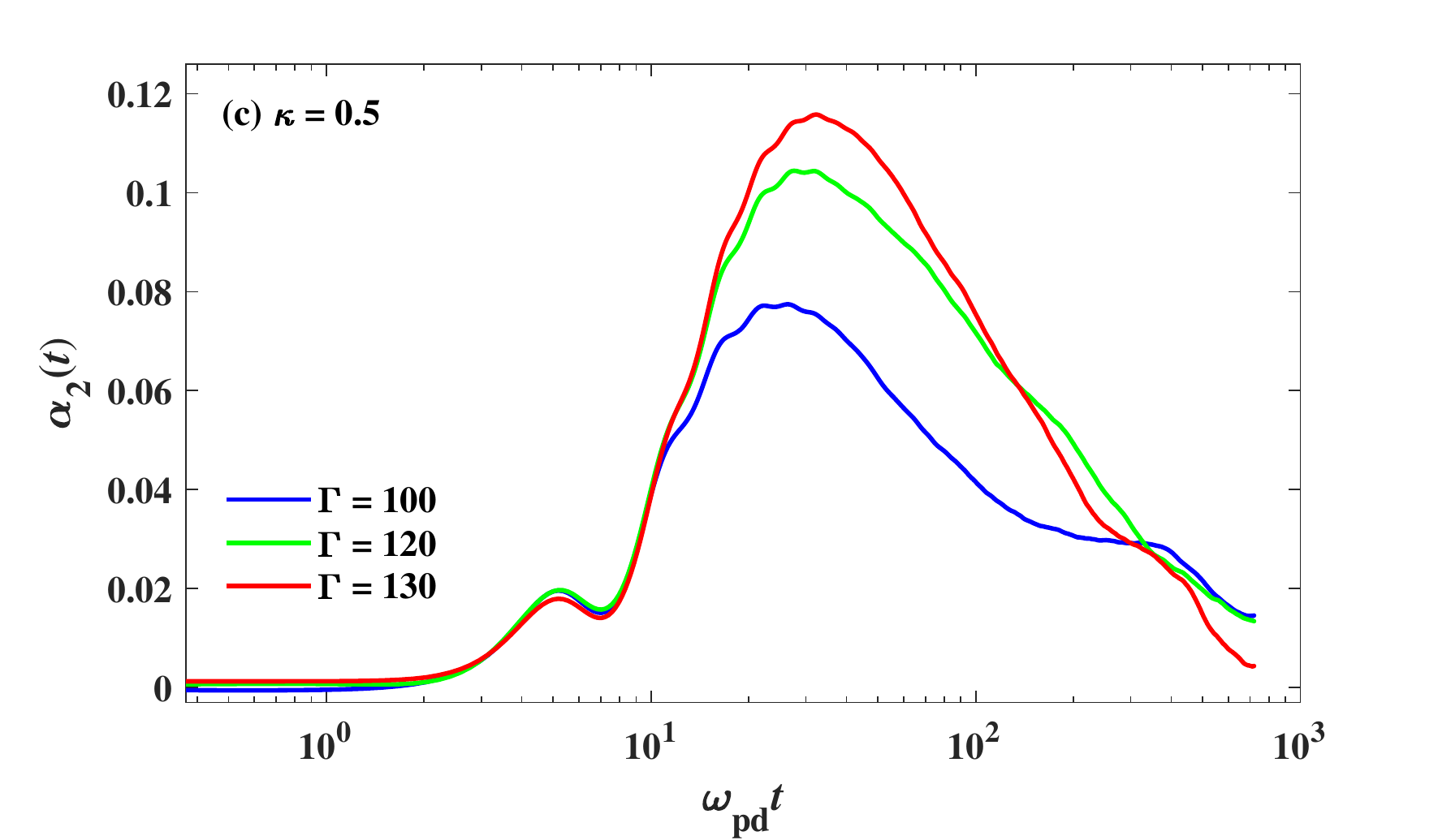}

\caption{Non-Gaussian parameter $\alpha_2(t)$ for different temperatures. (a) $\kappa=2.0$, (b) $\kappa=1.2$, and (c) $\kappa=0.5$. For each $\kappa$ value, the peak of $\alpha_2(t)$ increases with increasing $\Gamma$ or decreasing $T$.}\label{label8}
\end{figure}

\begin{figure*}[!ht]

\begin{minipage}[b]{.3\textwidth}
\includegraphics[width=5.0cm,height=5.5cm]{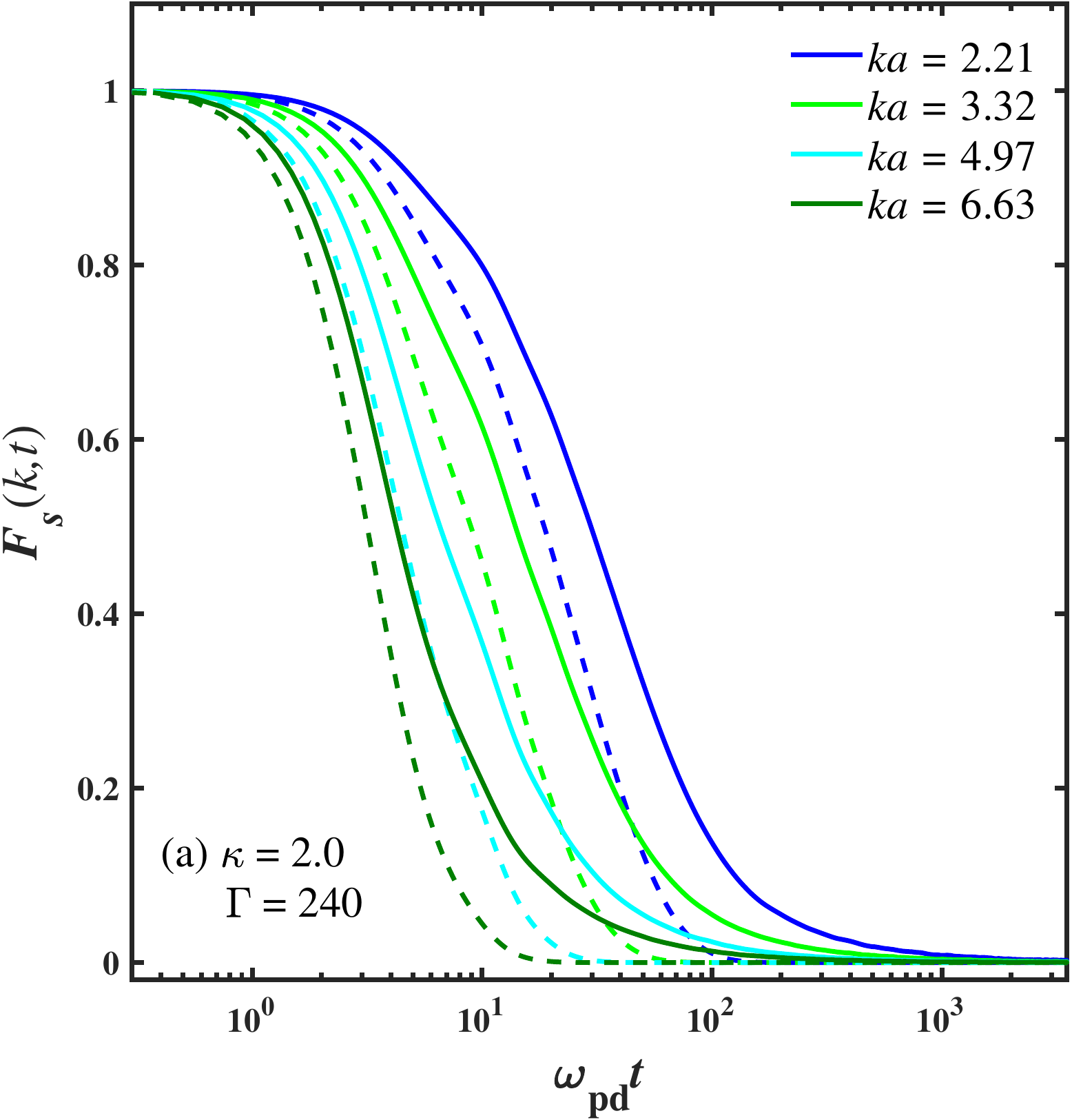}

\end{minipage}\qquad
\begin{minipage}[b]{.3\textwidth}
\includegraphics[width=5.0cm,height=5.5cm]{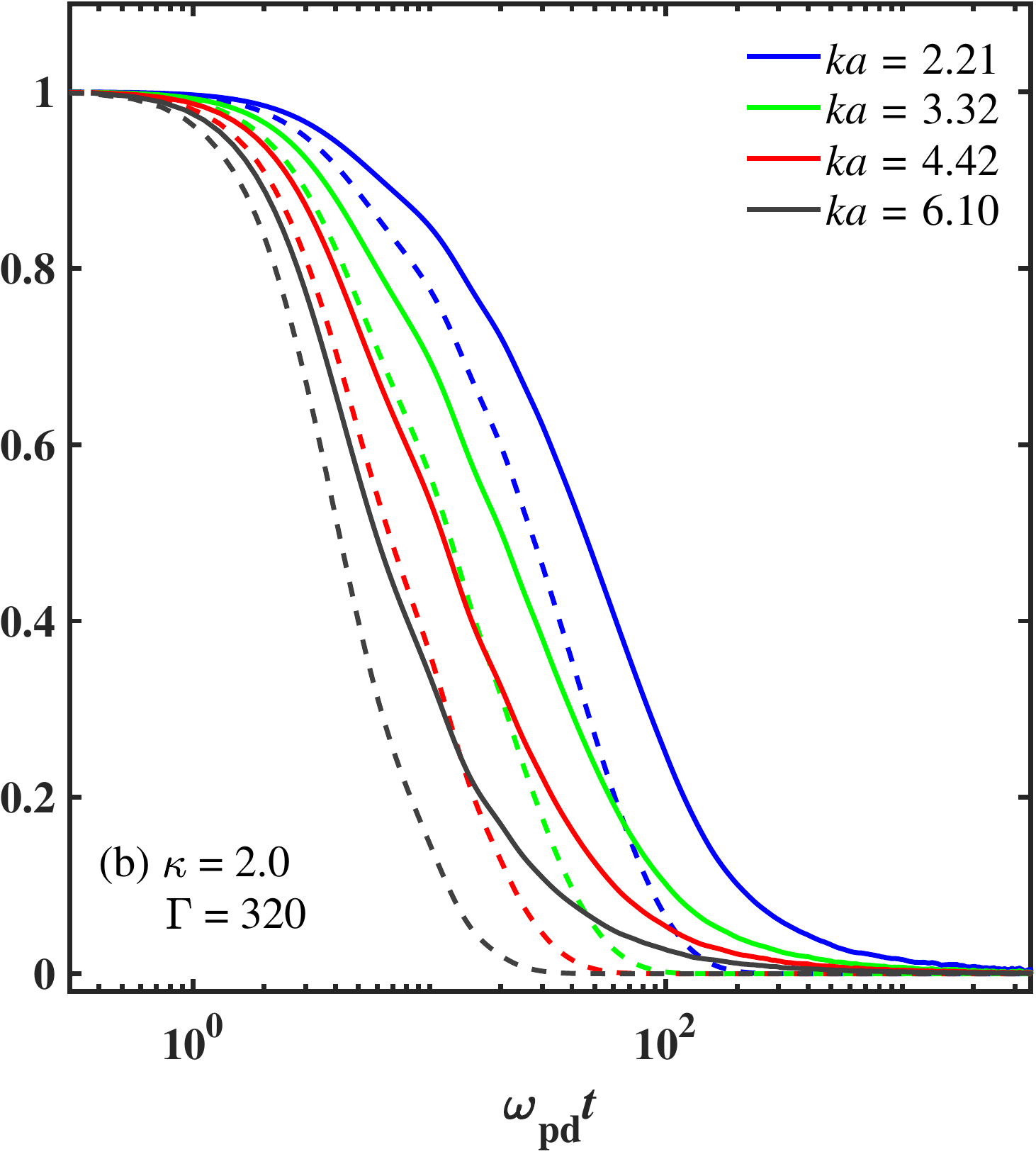}

\end{minipage}
\begin{minipage}[b]{.3\textwidth}
\includegraphics[width=5.0cm,height=5.5cm]{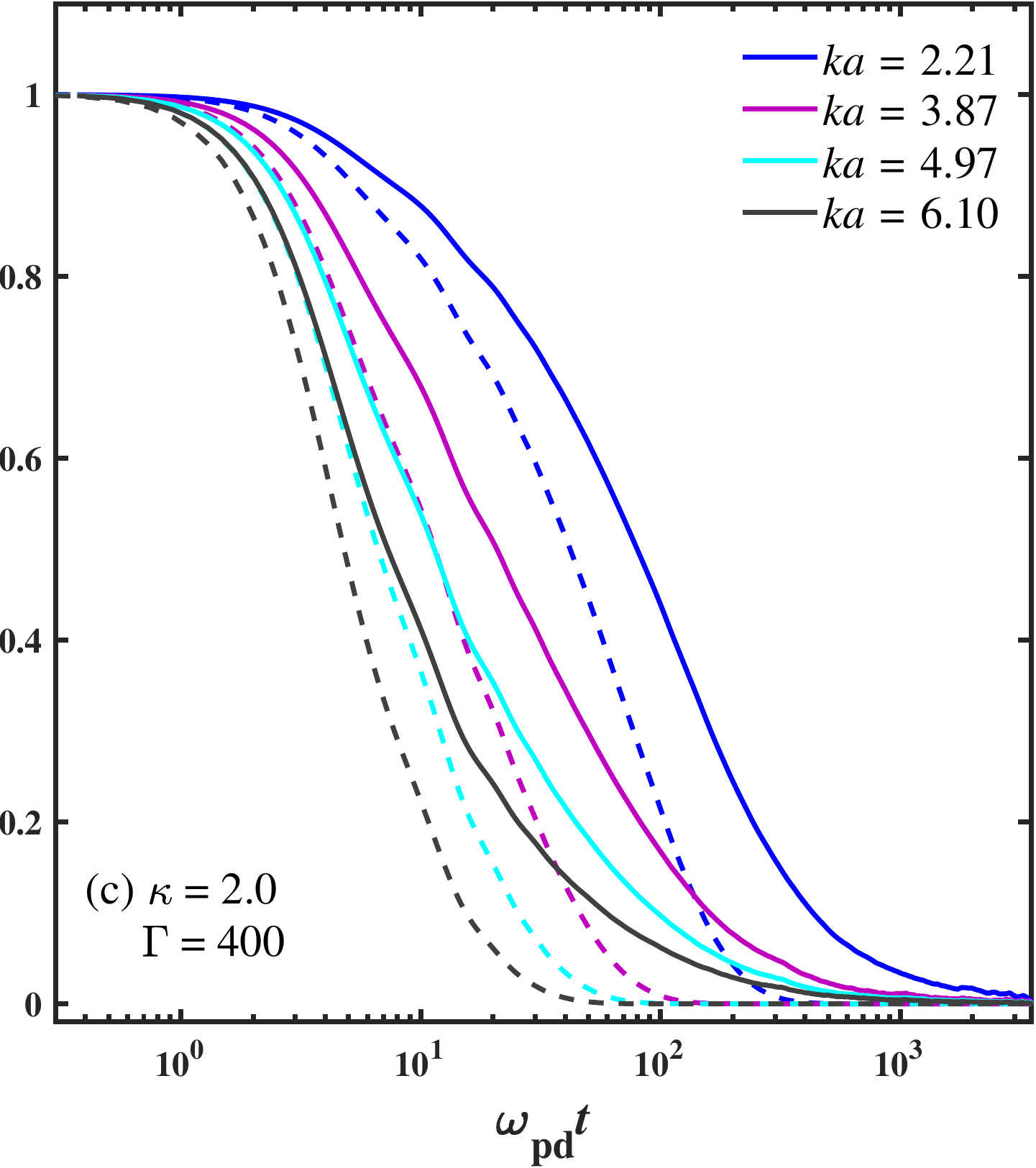}

\end{minipage}\qquad

\end{figure*}
\begin{figure*}[!ht]

\begin{minipage}[b]{.3\textwidth}
\includegraphics[width=5.0cm,height=5.5cm]{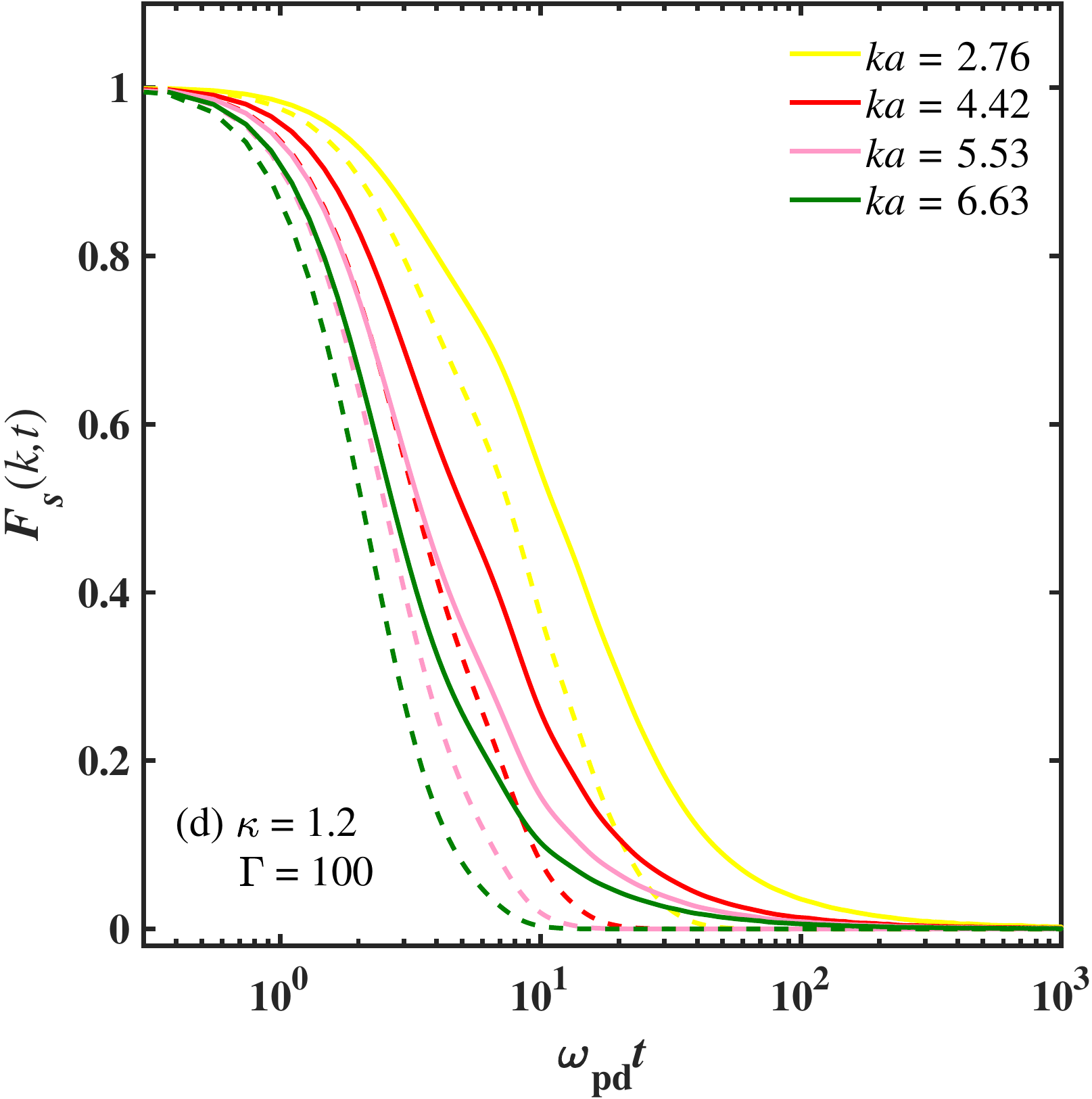}

\end{minipage}\qquad
\begin{minipage}[b]{.3\textwidth}
\includegraphics[width=5.0cm,height=5.5cm]{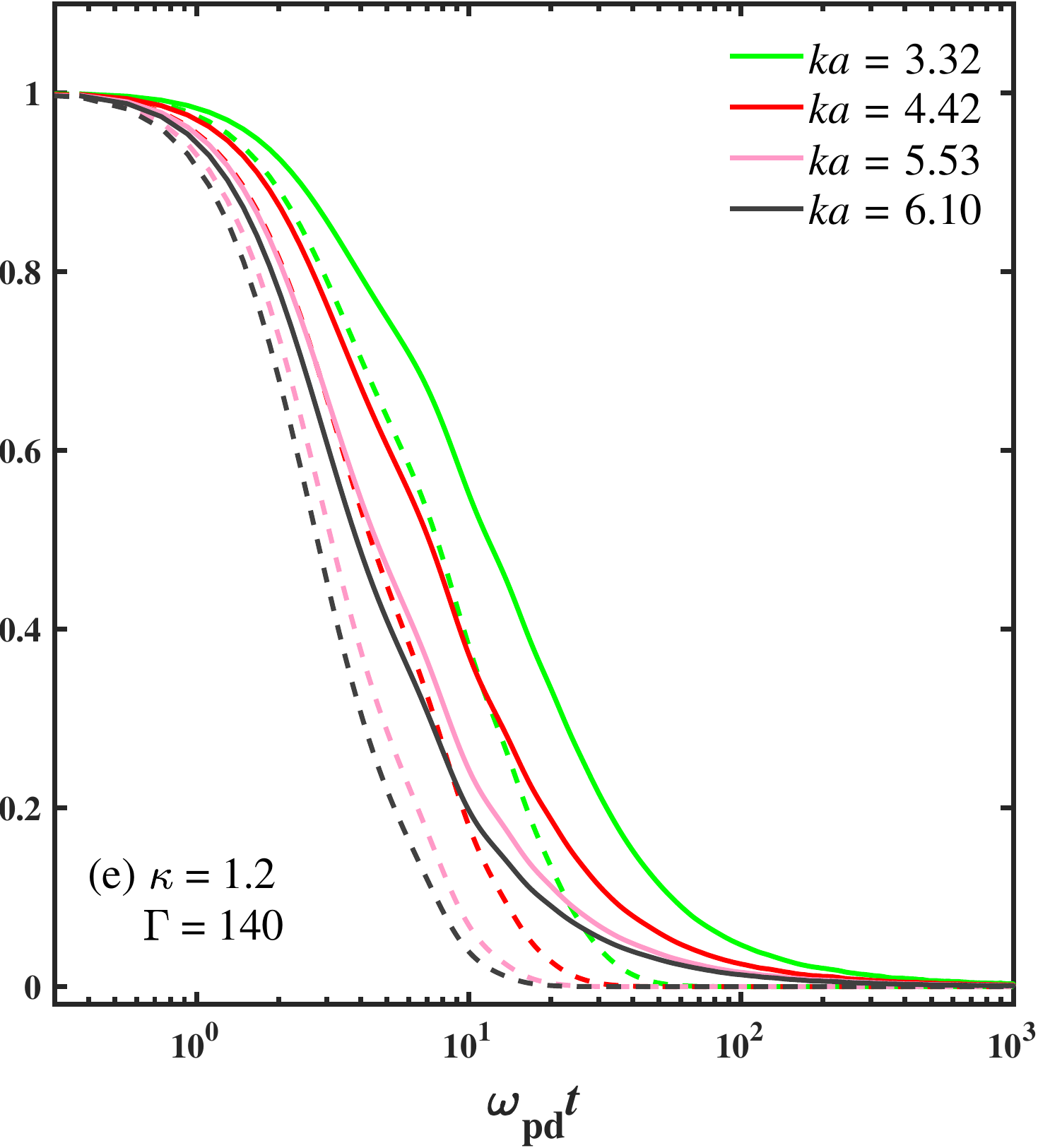}

\end{minipage}
\begin{minipage}[b]{.3\textwidth}
\includegraphics[width=5.0cm,height=5.5cm]{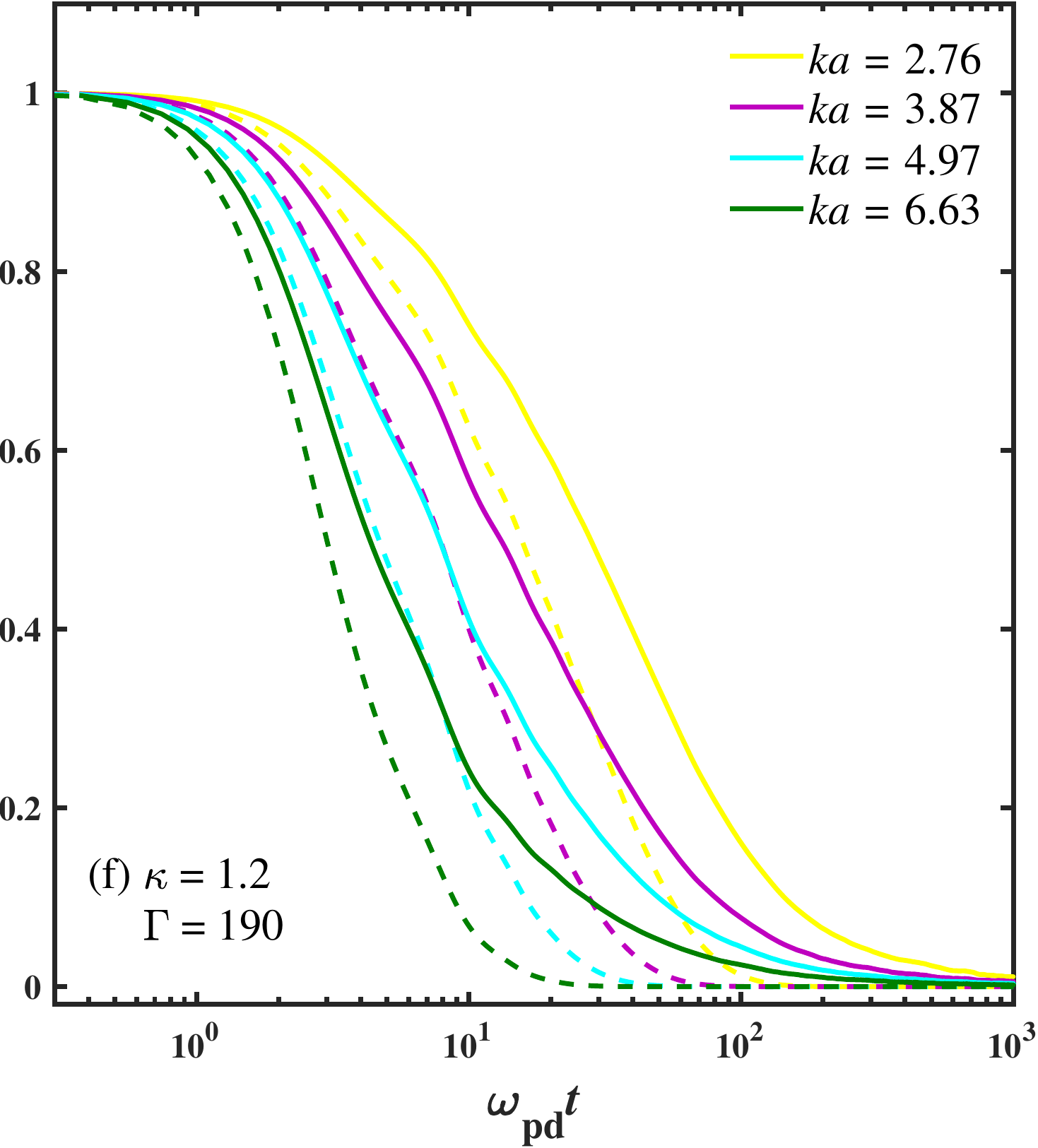}

\end{minipage}\qquad

\end{figure*}
\begin{figure*}[!ht]

\begin{minipage}[b]{.3\textwidth}
\includegraphics[width=5.0cm,height=5.5cm]{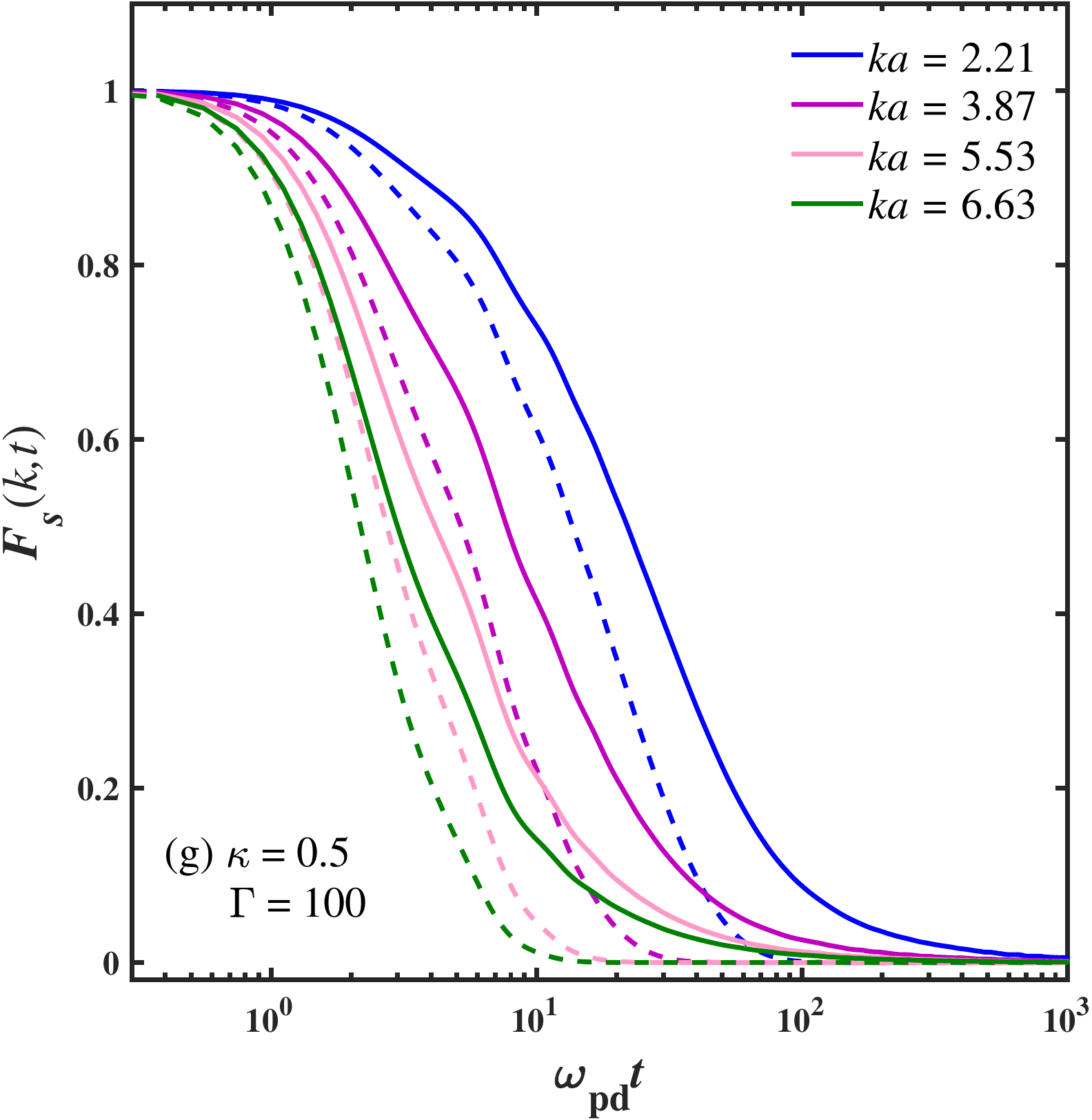}

\end{minipage}\qquad
\begin{minipage}[b]{.3\textwidth}
\includegraphics[width=5.0cm,height=5.5cm]{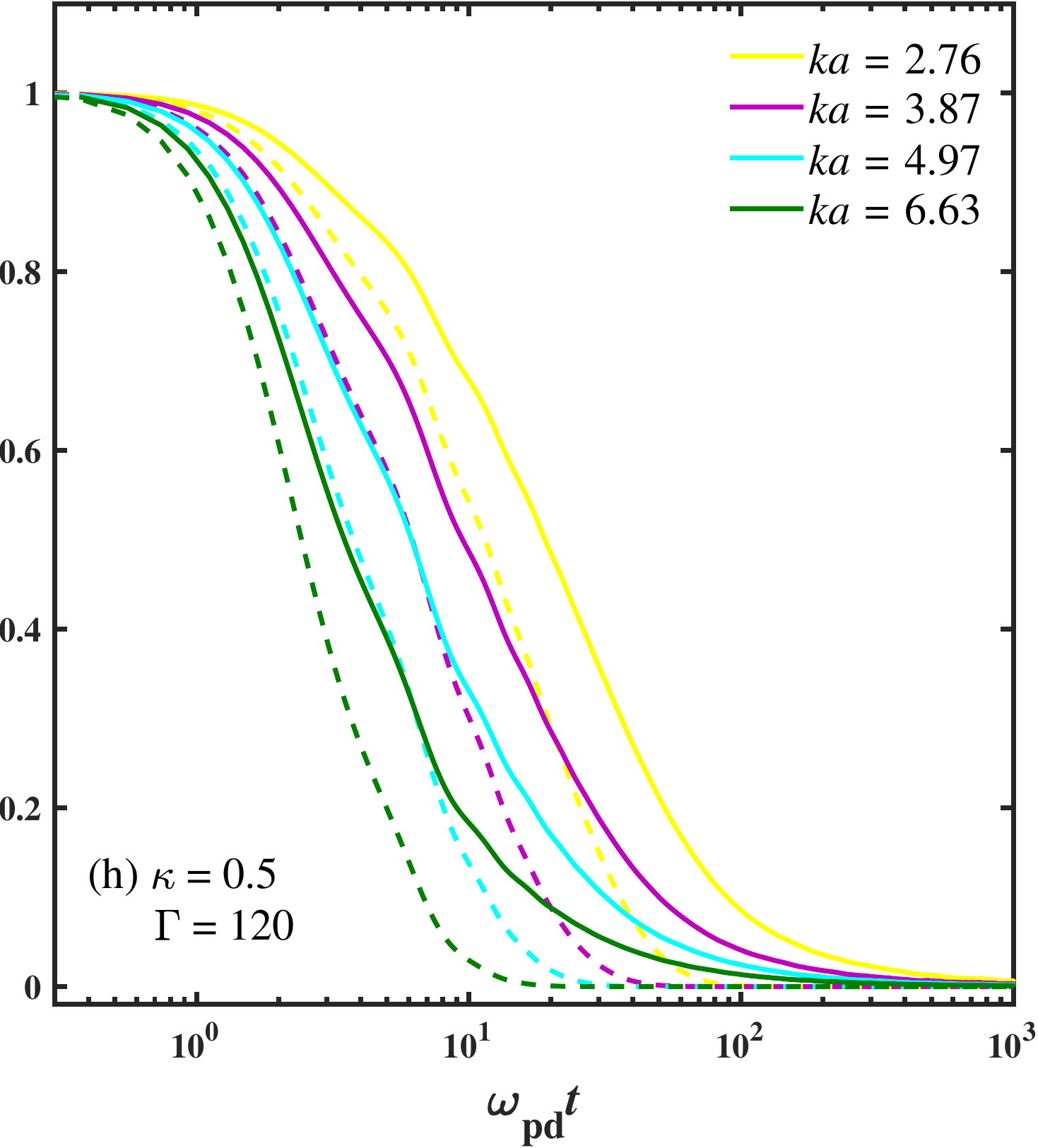}

\end{minipage}
\begin{minipage}[b]{.3\textwidth}
\includegraphics[width=5.0cm,height=5.5cm]{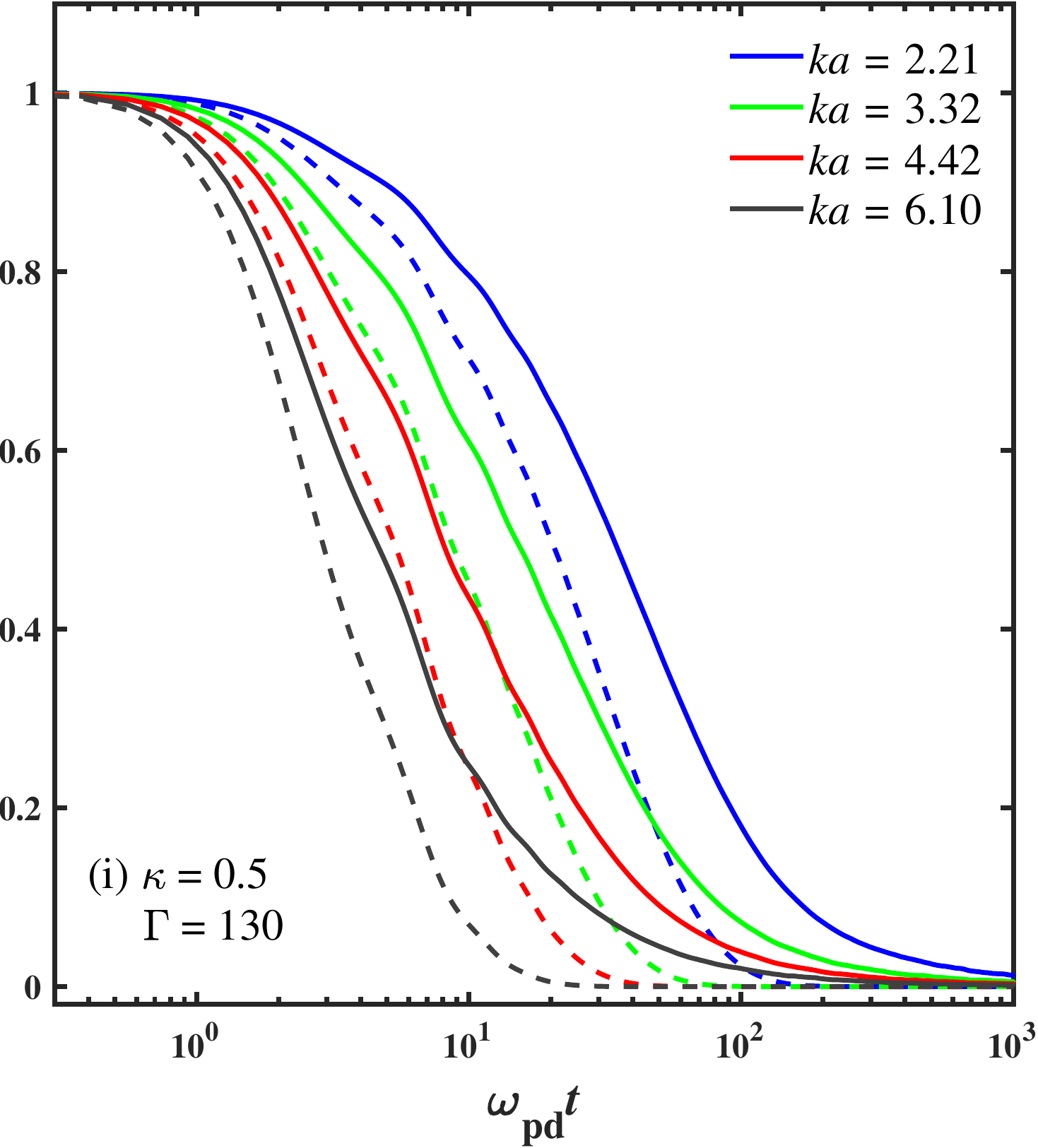}

\end{minipage}\qquad
\caption{Self-part of the intermediate scattering functions $F_s(k,t)$ for various dimensionless wave numbers $ka$ and the selected $\kappa$ and $\Gamma$ values. Solid lines are from the simulation, and dashed lines are from the Gaussian approximation obtained from Eq. (\ref{eq13}) with the MSD obtained from the simulation.}\label{label9}
\end{figure*}

\subsection{\label{SISF}Self-intermediate scattering function}

The quantity of interest in scattering experiments is the spatial Fourier transform of self-van Hove function, which is called the self-intermediate scattering function (self-ISF)~\cite{Hansen}
\begin{equation}
F_s(\boldsymbol{\mathrm{k}},t)=\int G_s(\boldsymbol{\mathrm{r}},t)\rm{exp}(-i\boldsymbol{\mathrm{k}}.\boldsymbol{\mathrm{r}})\mathit{d}\boldsymbol{\mathrm{r}},
\label{eq9}
\end{equation}
where $\boldsymbol{\mathrm{k}}$ is the 2D wave vector. In the MD simulation with periodic boundaries, the wave vectors are proportional to the periodicity of the system, i.e., $\boldsymbol{\mathrm{k}}= (2\pi/L)(k_x,k_y)$ where $L$ is the length of the simulation box and $k_x,k_y$ are integers~\cite{Allen}. The function $F_s(\boldsymbol{\mathrm{k}},t)$ is interpreted as the characteristic function of $G_s(\boldsymbol{\mathrm{r}},t)$ because according to the probability theory, the Fourier transform of a probability distribution function is called the characteristic function of the distribution~\cite{Berne1976}. Using the definition of the self-van Hove function from Eq. (\ref{eq4}) and applying the property of the delta function gives
\begin{equation}
F_s(\boldsymbol{\mathrm{k}},t)=\frac{1}{N}\Bigg\langle\sum_{\mathit{i}=1}^N \rm{exp}\left(-i\boldsymbol{\mathrm{k}}.[\boldsymbol{\mathrm{r}}_\mathit{i}(t)-\boldsymbol{\mathrm{r}}_\mathit{i}(0)]\right)\Bigg\rangle.
\label{eq10}
\end{equation}

For an isotropic system, $F_s(\boldsymbol{\mathrm{k}},t)$ depends only on the magnitude $k= \vert\boldsymbol{\mathrm{k}}\vert$, therefore, averaging over all directions yields 

\begin{equation}
\big\langle\rm{exp}\left(-i\boldsymbol{\mathrm{k}}.\boldsymbol{\mathrm{r}}\right)\big\rangle_{\phi}=\frac{1}{2\pi}\int_0^{2\pi}\rm{exp}\left(-i\mathit{kr} cos\phi\right)\mathit{d}\phi=\mathit{J}_0(\mathit{kr})
\label{eq11}
\end{equation}
where $\phi$ is the angle between the vectors $\boldsymbol{\mathrm{k}}$ and $\boldsymbol{\mathrm{r}}$, and $\mathit{J}_0(\mathit{kr})$ = sin($kr$)/($kr$) is the ordinary Bessel function of order zero. Therefore, for an isotropic system, $F_s(k,t)$ reduces to 
\begin{equation}
F_s(k,t)=\frac{1}{N}\Bigg\langle \sum_{\mathit{i}=1}^N\frac{\mathrm{sin}(k\vert \boldsymbol{\mathrm{r}}_\mathit{i}(t)-\boldsymbol{\mathrm{r}}_\mathit{i}(0)\vert)}{k\vert \boldsymbol{\mathrm{r}}_\mathit{i}(t)-\boldsymbol{\mathrm{r}}_\mathit{i}(0)\vert} \Bigg\rangle.
\label{eq12}
\end{equation}

We computed $F_s(k,t)$ from the equation (\ref{eq12}) for various wave numbers $k$ and the selected $\kappa$ and $\Gamma$ values. The results are shown in Figs. 9(a)--9(i).
If the system exhibits Fickian diffusion, the self-van Hove function is Gaussian at all times and its Fourier transform is obtained by substituting Eq. (\ref{eq2}) in Eq. (\ref{eq9}) as follows
\begin{equation}
F_s(k,t)=\mathrm{exp}\left(-\frac{k^2\langle(\Delta \mathrm{r}(t))^2\rangle}{4}\right),
\label{eq13}
\end{equation} \\
which is known as the Gaussian approximation~\cite{Hansen}. As shown in Figs. 9(a)--9(i), by comparing the self-ISF obtained by simulations and the corresponding Gaussian approximation, i.e., Eq.(\ref{eq13}), in which $\langle(\Delta \mathrm{r}(t))^2\rangle$ is determined from
the simulation data, we find that at intermediate times, the Gaussian approximation fails to describe the diffusion of dust particles, indicating the non-Gaussian dynamics in 2D Yukawa liquids. At short and very long times, as we expected the Gaussian approximation is a good description for the self-intermediate scattering function.

The origin of this non-Gaussian behavior may be heterogeneous dynamics, which has been observed in liquids~\cite{Reichhardt2003,Kob1997}.
Dynamical heterogeneity reflects the existence of regions in which dust particles are more mobile than expected from a Gaussian approximation, that is, dust particles that move faster than the rest. These dust particles form clusters, i.e., groups of the particles and move along stringlike (one dimensional) paths. Therefore, strings of mobile dust particles flow among regions including less mobile dust particles~\cite{Liu2006}. As a result, the displacement deviations appear over time and the distribution function of the particle displacement, i.e., $G_s$ departs from Gaussian shape. Consequently, $\alpha_2(t)$ is not zero.

\section{\label{Conclusions}SUMMARY AND CONCLUSIONS}
We have investigated the dynamics of dust particles in two-dimensional Yukawa liquids using molecular dynamics simulation. First, we have computed the mean-squared displacement on the allowed ranges of experimental parameters and have shown that it is linear with time (Fickian diffusion) in very low temperatures, on the time scales which dust particles diffuse. Then, we have computed the distribution of the particles displacements $G_s(r,t)$, and have compared it with $G_s(r,t)$ obtained from Gaussian approximation. Significantly, we found that at intermediate times, the distribution of the particles displacements deviates from the Gaussian, i.e., the failure of the Gaussian approximation, which states that when diffusion is Fickian, the distribution of particles displacements is Gaussian. This result may be attributed to the heterogeneous dynamics of dust particles in Yukawa liquids. Non-Gaussian parameter and self-intermediate scattering function have also been computed and their results have confirmed these deviations. Furthermore, we found that the deviations increase with decreasing the temperature of the liquid.  
Here, we have decreased the temperature to near the melting point $T_m$, where the liquid phase is maintained.
A further decrease in temperature results in a phase change of the system from liquid to solid.
Investigating the non-Gaussian behavior and the degree of deviation from Gaussian in Yukawa solids can be an interesting research topic. Our future work will include this research.

\nocite{*}

\bibliography{Ghannad}

\end{document}